\documentclass[emulateapj5]{aastex}

\usepackage{apjfonts}
\usepackage{emulateapj5}
\usepackage{psfig}

\def\erg/cm2sec{ergs~cm$^{-2}$~s$^{-1}$}  
\def\ergcm2{ergs~cm$^{-2}$}

\newcommand{\lsim }{{\lower0.8ex\hbox{$\buildrel <\over\sim$}}}
\newcommand{\gsim }{{\lower0.8ex\hbox{$\buildrel >\over\sim$}}}

\def\apj{ ApJ}
\def\aap{ A\&A}

\def\Chandra{${\it Chandra}$\ }
\def\HST{${\it HST}$\ }
\def\RXTE{${\it RXTE}$\ }

\def\simge{\mathrel{%
   \rlap{\raise 0.511ex \hbox{$>$}}{\lower 0.511ex \hbox{$\sim$}}}}
\def\simle{\mathrel{
   \rlap{\raise 0.511ex \hbox{$<$}}{\lower 0.511ex \hbox{$\sim$}}}}

\newcommand{\Msun}{\ifmmode {M_{\odot}}\else${M_{\odot}}$\fi}
\newcommand{\Rsun}{\ifmmode {R_{\odot}}\else${R_{\odot}}$\fi}

\shorttitle{Terzan 5}
\shortauthors{Heinke et al.}

\begin{document}
\title{A \Chandra X-ray Study of the Dense Globular Cluster Terzan 5}   

\author{C. O. Heinke,  P. D. Edmonds, J. E. Grindlay, D. A. Lloyd}
\affil{Harvard College Observatory,
60 Garden Street, Cambridge, MA  02138;\\
 cheinke@cfa.harvard.edu, pedmonds@cfa.harvard.edu,
josh@cfa.harvard.edu,  dlloyd@cfa.harvard.edu}
\and
\author{H. N. Cohn and P. M. Lugger}
\affil{Department of Astronomy, Indiana University, Swain West 319,
Bloomington, IN 47405; cohn@indiana.edu, lugger@indiana.edu}

\begin{abstract}
We report a \Chandra ACIS-I observation of the dense globular cluster
Terzan 5.  The previously known transient low-mass x-ray binary (LMXB)
EXO 1745-248 in the cluster 
entered a rare high state during our August 2000 observation, complicating the
analysis.  Nevertheless nine additional sources clearly associated
with the cluster are also detected, ranging from $L_X(0.5-2.5{\rm
keV})=5.6\times10^{32}$ 
down to $8.6\times10^{31}$ ergs s$^{-1}$.  Their X-ray colors and
luminosities, and spectral fitting, indicate that five 
of them are probably cataclysmic variables, and four are likely  
quiescent LMXBs containing neutron stars.   We
estimate the total number of sources between $L_X(0.5-2.5 {\rm
keV})=10^{32}$ and $10^{33}$ ergs s$^{-1}$ as $11.4^{+4.7}_{-1.8}$ by
the use of artificial point source 
tests, and note that the numbers of X-ray sources are similar to 
those detected in NGC 6440.  
The improved X-ray position allowed us to identify a plausible
infrared counterpart to EXO 1745-248 on our 1998 {\it Hubble Space Telescope}
NICMOS images. This blue star
(F110W=18.48, F187W=17.30) lies within 0.2'' of the boresighted LMXB
position.   Simultaneous {\it Rossi X-ray Timing Explorer} (\RXTE) spectra,
combined with the \Chandra spectrum, indicate that EXO 1745-248 is
an ultracompact binary system, and show a strong broad 6.55 keV iron
line and an 8 keV smeared reflection edge.

\end{abstract}


\maketitle

\section{Introduction}

The high resolution of the \Chandra {\it X-ray Observatory}  has enabled
astronomers to study the low-luminosity X-ray source populations in
globular clusters in great detail.  Combined X-ray, radio, and optical
{\it Hubble Space Telescope} (\HST) studies of the globular cluster 47
Tucanae have 
revealed quiescent low-mass X-ray binaries (qLMXBs) that have not
experienced X-ray outbursts in the history of X-ray astronomy,
cataclysmic 
variables (CVs) as X-ray luminous as any known in the field,  flaring
behavior from coronally active stellar binary systems, and predominantly 
thermal X-ray emission from millisecond pulsars (MSPs) (Grindlay et
al. 2001a, 2002).  Similar populations have been
uncovered in the globular clusters NGC 6397 (Grindlay et al.~2001b),
NGC 6752 (Pooley et al.~2002a), and $\omega$ Cen (Rutledge et
al. 2001, Cool, Haggard \& Carlin 2002), while the luminosities and broad
X-ray spectral 
types of these sources have allowed classification of sources in the
more obscured cluster NGC 6440 (Pooley et al.~2002b) and M28 (Becker
et al.~2003).  

The globular
cluster Terzan 5 contains a known transient LMXB, EXO 1745-248, which
was first detected in a bursting state 
by {\it Hakucho} (Makishima et al.~1981), and has been irregularly 
active since then (Johnston et al.~1995 and refs. therein).  EXO
1745-248 is one of the few luminous globular cluster LMXBs not analyzed by
Sidoli et al.~(2001; hereafter 
SPO01) or Parmar et al.~(2001), who identify spectral
distinctions between normal and ultracompact LMXBs.  Ultracompact
LMXBs, defined as having periods less than 1 hour, are thought to
possess a degenerate helium white dwarf secondary.  Deutsch et
al.~(2000) remark upon the overabundance of ultracompact LMXBs in
globular clusters, and speculate that the short periods may be due to
dynamical effects in globular clusters.

Terzan 5 also contains two identified MSPs, with many additional MSPs
probably making up the extended steep-spectrum radio source at the
cluster core (Lyne et
al. 2000; Fruchter \& Goss 2000).  Terzan 5's high central density and large
mass make it a rich target for studies of binary systems, but its
high reddening and severe crowding make optical and even infrared
observations extremely difficult. 
The deepest infrared survey of Terzan 5 was performed with the \HST NICMOS
camera by Cohn et al.~(2002; hereafter CLG02).  The extreme reddening
indicates that the cluster parameters are best determined in the
infrared.  NICMOS observations produced the first CMDs of Terzan 5 to reach the
main-sequence turnoff (CLG02; see also Ortolani et al.~2001).
Therefore we utilize the new cluster 
parameters derived by CLG02 in our analysis, particularly the
reddening, distance, core radius, and radial star-count profiles.  The 
attempt by Edmonds et al.~(2001, hereafter EGC01) to identify EXO
1745-248 and the eclipsing MSP through time-series variability analysis and
color information did not
uncover any promising candidates due to the crowding and overlapping
Airy profiles, although it did identify two truly
variable stars, one of which was shown to be an RR Lyrae variable.

This paper is organized as follows.  Section 2.1 describes the
observations we used.  Section 2.2 explains our methods for detecting
sources and performing an astrometric correction based upon
identification of serendipitous sources. Section 2.3 describes our
search for an infrared counterpart for EXO 1745-248, extending the
work of EGC01 using our \Chandra position.  Section 2.4 classifies the
faint X-ray sources, and 
quantifies our detection incompleteness (due to the outburst of EXO
1745-248). Section 2.5 attempts simple spectral fits to the
faint sources, while section 2.6 examines the simultaneous \RXTE and
\Chandra spectra of EXO 1745-248.  Section 3.1 compares our spectral
analysis of EXO 1745-248 to other observations of LMXBs.  Section 3.2
recalculates the central 
density and collision frequency of Terzan 5, while section 4
summarizes our conclusions. 

\section{Analysis}

\subsection{Description of Observations}

The \Chandra X-ray Observatory observed the globular cluster
Terzan 5 on July 24, 2000, for 45 ksec (05:46 to 18:22 TT), and on
July 29, 2000 for 5 ksec (00:56 to 09:20 TT),
with the ACIS-I instrument at the focus.  Due to an error in the
observation upload, both exposures were performed in 1/8 subarray
mode (as intended for the shorter observation only), with frame times
of 0.841 s (longer exposure with more chips) 
and 0.541 s.  EXO 1745-248 entered an outburst  
during July 2000, its 2-10 keV flux varying between 54 and 600 mCrab during
July and August, approaching its Eddington limit at maximum flux (Markwardt et
al. 2000a, 2000b).  {\it Rossi X-ray Timing Explorer} (\RXTE) All-Sky
Monitor (ASM) observations (results provided
by the ASM/\RXTE team\footnote{See http://xte.mit.edu/.})
show that the 2-10 keV countrate from EXO 1745-248 on July 24, 2000,
was  5.3$\pm0.9$ cts s$^{-1}$ ($\sim$ 72 mCrab) 
during our longer \Chandra observation, rising to 9.8$\pm0.9$ cts
s$^{-1}$ in the second observation (Fig. 1).  

Unfortunately, the \Chandra observation of EXO 1745-248 was not
 optimized to study such a bright object (Fig. 2).  The intense photon
 flux led to 
severe pileup, which occurs when two or more photons landing in the
same or adjacent pixels between frames are recorded as a single
event.  Pileup can increase the energy of recorded events, or by
altering the grade of the recorded event to a ``bad'' grade, cause
the rejection of the event either before or after telemetry to the
 ground (See the \Chandra Proposer's Observatory
Guide, v.5, chapter 6). In this case the pileup was severe enough to
 cause the pre-telemetry rejection of all events recorded within
 $\sim$1'' of the LMXB 
 position. The LMXB's X-ray halo, by increasing the local
 background, greatly degraded our sensitivity to faint cluster
 sources. (The halo is due to a combination of dust grain scattering
 and the intrinsic breadth of the \Chandra mirrors' point spread
 function.) 
Nevertheless, the spectacular resolution of \Chandra did allow us
to identify additional point sources up to $10^5$ times fainter within
10'' of EXO 1745-248.  The readout streak (caused by photons from EXO
 1745-248 arriving during the readout of the CCD) also degraded our
 survey, but we were able to extract a useful spectrum of EXO 1745-248
 from the readout streak.   
We also analyzed a simultaneous {\it Rossi X-ray
Timing Explorer} (\RXTE) pointed observation from the HEASARC
 archive\footnote{Available at
 http://heasarc.gsfc.nasa.gov/docs/corp/data.html.} 
(on July 24, 2000, 15:15 to 16:16 TT), for
broad spectral coverage of the outburst of EXO 1745-248.  The \HST
NICMOS data we used to search for a 
possible infrared counterpart to EXO 1745-248 are described in EGC01
and CLG02.

\subsection{Detection and astrometry}

 We used the CIAO 
software package\footnote{Available at http://asc.harvard.edu/ciao/.}
to search for point sources, produce hardness ratios, and extract
spectra and lightcurves.  We
reprocessed the two observations to remove the pixel randomization
added in standard processing, and merged the two observations.  No
periods of high background flaring were observed. EXO 1745-248's X-ray
halo displays an x-ray color (defined here as 2.5 log [0.5-1.5 keV 
counts/1.5-6 keV counts], following Grindlay et al. (2001a)) of -2.9,
harder than most known faint globular cluster sources.  Therefore, we
selected a soft band, 0.5-2 keV, to search for point sources.  We
selected a $2.5\times10^4$ pixel (1.7 arcmin$^2$) square region including the 
cluster and ran WAVDETECT with the significance threshold set to give
false positives at the rate of $10^{-4}$.  WAVDETECT indeed found two
 spurious sources (identified as spurious by eye, and by not appearing
in more than one energy band) far from the
cluster, as expected.  Within three optical core radii (24'') we
identify nine point  
sources besides EXO 1745-248, each confirmed by visual inspection (see 
Figure 2) and with significance above 2.8 $\sigma$.    We name the
nine additional real sources in Table 1 with both positional 
names and simple reference names W2-W10 (used in the rest of this paper).
Outside the globular cluster, we binned the remaining data into 1 
arcsecond pixels and searched for back- or foreground serendipitous
sources (setting the significance threshold to a false rate of
$10^{-6}$).   We 
identified four sources that also pass visual inspection on the active
portions (36 arcminutes$^2$) of the ACIS-I array.  As they are all
more than 2' (2.5 half-mass radii) from the cluster center, we
hereafter identify them as ``serendipitous sources''.  The ROSAT 
source S2 from Johnston et al.~(1995) was not 
included in the field of view.  Inspection of the locations of the
known millisecond pulsars (A and C; Lyne et al.~2000) show no evidence of X-ray
emission (using the astrometry below).  The upper limits are 3 and 5
counts respectively in the 
0.5-2 keV band, giving $L_X\leq2\times10^{31}$ and $\leq3\times10^{31}$
ergs s$^{-1}$ for a power law of photon index 2.  These are well above
the X-ray 
luminosities of all identified millisecond pulsars in 47 Tuc (Grindlay et
al.~2002), NGC 6397 (Grindlay et al.~2001b), and NGC 6752 (D'Amico et
al. 2002), but below the luminosity ($L_X=1.1\times10^{33}$ ergs
s$^{-1}$) of the MSP B1821-24 in M28 (Danner et al.~1997).  Soft
thermal spectra give even less constraining limits, 
due to the high extinction.  The two variables
identified by EGC01 are also not seen, as expected
based on their RR Lyrae and probable eclipsing blue straggler
identifications.  

Inspection of a Digital Sky Survey image reveals that one of
the serendipitous sources has a probable optical counterpart.
According to the Guide Star Catalog 2.2, a star with V=13.56, R=13.22 is
located only 0\farcs36 from the position derived by 
WAVDETECT for the X-ray source that we name CXOU J174803.3-244854.
This is consistent with the 0\farcs6 absolute astrometry (90\%
conf. radius) reported by Aldcroft et al. (2000) for \Chandra.  
We estimate a probability of a positional coincidence within 0\farcs5 of
one of the 15 brightest stars on the 7' by 7' survey plate with any of
the four serendipitous X-ray 
sources of $8\times10^{-5}$.  Another X-ray
source (CXOU J174812.6-244811.1) has a faint star (R=16.8) 0\farcs14 away
(when the frame is shifted to match star 1).  We estimate the chances
of a star this bright or brighter landing within 0\farcs5 of one of
four \Chandra sources as $4\times10^{-3}$.   The other
two X-ray sources show no stars in the GSC 2.2 Catalog within 3''.  We
 shift our \Chandra frame by
+0\fs004 in RA, and +0\farcs396 in Dec, to match the weighted GSC 2.2 star
positions.\footnote{GSC 2.2 absolute astrometric errors, when compared
to the international celestial reference frame, are of order 0.3-0.75
arcseconds; see http://www-gsss.stsci.edu/gsc/gsc2/GSC2home.htm.}

The V and R magnitudes of star 1 
suggest an F star, and the X-ray (0.5-2.5 keV) to V-band flux ratio
(V-band flux 
defined as $10^{-0.4V-5.43}$ ergs cm$^{-2}$ s$^{-1}$) is
$2\times10^{-4}$, on the high end of values for nearby F stars
(H\"unsch et al.~1999).  Star 2 has no color information in the GSC
2.2 Catalog.  Assuming
V-R$\sim$1.5 (appropriate for M2 stars) gives a flux ratio of
$1\times10^{-2}$, which is common among M stars (H\"unsch et
al. 1999).   The other two sources may be background AGN, or 
(perhaps more likely) CVs in the galactic bulge.  Both show 
an Xcolor (2.5 log [0.5-1.5 keV cts/1.5-6 keV cts]) of -1.5,  
indicating strong absorption. Simple spectral fitting of the brighter
source gives $N_H=2.8^{+2.4}_{-1.2}\times10^{22}$ cm$^{-2}$ for a
powerlaw of photon index 1.7, as typical for AGN.  A 10 keV
bremsstrahlung spectrum (as for a bulge CV) would have
$N_H=2.7^{+2.5}_{-1.2}\times10^{22}$ cm$^{-2}$.  The 2-10 keV log
N--log S relation of Giacconi et al. (2001) suggests that 1-3
AGN may be expected at these observed flux levels in our ACIS-I
subarray field (although the location of detected sources near to the
aimpoint suggests that our sensitivity is not uniform across the
field).   Recent results from galactic bulge surveys (Grindlay
et al.~2003) indicate that bulge CVs or compact binaries outnumber AGN
at these flux levels, suggesting that these two sources may be bulge CVs.

We found the position of EXO 1745-248 by centering the 
symmetric ``hole'' caused by extreme pileup and the
readout streak.  We estimate our error in this determination at 2/5 of
a pixel, or 0\farcs2.
We present the positions (in the GSC 2.2 frame) of EXO 1745-248, the nine
additional globular 
cluster sources, and the four serendipitous sources (with relative
positional errors from WAVDETECT) in Table 1, along
with the background-subtracted counts in three bands.

\subsection{Identification of plausible infrared counterpart to EXO 1745-248}

Our refined position for EXO 1745-248 allowed us to undertake a careful
search of the small \Chandra error circle in the June 1998 {\it Hubble Space
Telescope} NIC2 F110W and F187W data, for which the photometric
analysis is described in EGC01.  We 
estimate an error of 0\farcs1 in the shift to the GSC 2.2 frame, based
on the scatter of the two stars used in the boresighting.  An
additional error of 0\farcs37 (following EGC01) is incurred in the shift
between the 
GSC 2.2 and \HST NICMOS frames.  Combining these with the 0\farcs2
centering 
positional uncertainty gives an error for the position in the
NICMOS frame of 0\farcs4.  This error circle is fortuitously free of
bright red giants.  Error circles for six other X-ray sources in the
cluster lie wholly or partially within the NIC2 field of view.

Comparison of the F110W and F187W images with a ratio image (Figure
3) shows a significantly blue faint star (hereafter star A) within
the \Chandra error circle, only 0\farcs2 from the corrected position
of EXO 1745-248.  A combination of 
this star and another star just above it (Star B) was
shown in the cleaned CMD of EGC01 (their figure 8) as 
one of only two stars that lie more than 3$\sigma$ to the blue from the
distribution of stars in that color-magnitude diagram.  
Our attempts at separate PSF-fitting using ALLSTAR in
DAOPHOT encountered  two problems: (1) we were 
unable to fit stars A and B simultaneously and (2) the sky-fitting
component in PHOT and ALLSTAR gave local sky (background) values that were too
high, because stars A and B are surrounded by a ring of bright stars.
The
latter is a formidable problem in this crowded field, especially in F187W
where the background contribution of neighboring red giants is greater
than in F110W.  

We therefore adopted a customized procedure to estimate the F110W and F187W
magnitudes for stars A and B. For sky determination we adopted the mean of
the two 2-by-2 pixel regions having the lowest intensity values within an 8
pixel radius of star B.  Using the F110W and F187W images, and the ratio
image, we determined the positions of stars A and B. Using PSF models
and the DAOPHOT program ADDSTAR to create models for stars A and B we
subtracted these models from the two images to produce residual images,
while keeping the positions of stars A and B fixed. Since no other
close-by stars are obviously present, stars A and B are accurately
modeled if the mean aperture counts (after subtracting the sky) at their
positions in the residual image are equal to zero. We iteratively varied
the magnitudes of these two stars until this goal was met.
The results of these fits are shown in the lower right-hand corner of
Figure 3.   Clearly the subtractions of stars
A and B were successful, given the relatively smooth residuals.  

We performed aperture photometry on the residual images to produce
photometry that is consistent with Figure 8 of EGC01. Photometry for
stars A and B were measured by subtracting off 
stars B and A respectively, and the results are shown in Figure
4. Variables 1 and 2 from EGC01 are labeled 
with crosses, and stars within 1" of X-ray sources are circled.  The
main-sequence turnoff (not reached at the cluster core) is at
F110W$\approx20$ (CLG02).  Note the
apparently blue color of star A ($\sim2.5\sigma$ from the subgiant
sequence) lying  
near the top of the blue straggler sequence. The error in the color of
star A is dominated by the 
error in the F187W magnitude.  The high sky background in F187W
caused the original F187W magnitude determination for the combination of stars
A and B to be too low, and thus shifted the combination bluewards. 
Our customized procedure thus gives redder colors for both stars A and
B than the original determination for the combination.  

Given the blue appearance of star A in both the normal and residual images,
the reasonable evidence for blue colors in our photometric analysis, and
its proximity to the center of the X-ray error circle, star A is a
reasonable candidate for the IR counterpart to EXO 1745-248. No evidence is
seen for statistically significant variability of star A, but the faintness
of star A and its strong blending with star B limits the depth of
variability searches. Only the F110W images are useful for time series
analysis, and only sparse coverage was obtained here over 1.16 days.
Images in the $V$ and $I$ passbands with the Advanced Camera (in High
Resolution mode) on \HST should do a
much better job of separating stars A and B and would provide much better
constraints on variability if taken at several different epochs near
outburst and quiescence.

Since the F187W band does not correspond to any ground-based filter,
standard calibrations for it do not exist.  We apply rough
transformations to estimate interesting quantities.
We estimate the unreddened J magnitude from the F110W and F187W
magnitudes by the equation 

$J_0=F110W+0.033-0.178(F110W-F187W)+A_J$

\noindent 
derived assuming that the spectral slope is constant over this range.   
Using $A_J/A_V$=0.282 and $A_V$=6.75 (CLG02), we find
$J_0$=16.40, $M_J$=1.7.  We calculate the color
(F110W-F187W)$_0$= $-0.17\pm0.3$.  
Assuming that the photometry is accurate, three alternatives to an
LMXB nature remain for this star: it could be a blue 
horizontal branch star, it could be a blue straggler, or it could be a
foreground star.  We plot simple interpolated blue straggler
and horizontal branch sequences in Figure 4.  Star A's
position may be consistent with that of a blue horizontal branch star,
although a blue horizontal branch has only been hinted at in Terzan 5
(CLG02).  It is also consistent with an extremely heavy (more than twice the
turnoff mass) blue straggler, or a blue straggler binary (as variable 2
from EGC01, plotted as an ``X'' next to star A in figure 4, may be).  The other
extremely blue star seen in 
Figure 4 is probably a foreground star, which also cannot be
excluded as a possibility for star A.  However, the chance of
one of the five bluest stars in the 19\farcs2 square NIC2 FOV landing
in the 0\farcs4 error circle by chance is only $7\times10^{-3}$.  
None of the stars in the other X-ray error circles (circled in Figure
4) display any unusual
colors or variability.  This is expected based upon the magnitudes of CVs and
quiescent LMXBs, in the
field and other globular clusters, which very rarely reach the turnoff
magnitude (Grindlay et al.~2001a, b, Pooley et al.~2002a).

We compare this suggested counterpart to other low-mass X-ray binaries
observed in the infrared by calculating the ratio of X-ray to
infrared luminosity.  Using the ASM lightcurve in figure 1, we see
that the average ASM countrate for the 11 days leading up to the \HST
observation was $1.5\pm0.2$ cts s$^{-1}$, compared to 5.3$\pm0.9$
cts s$^{-1}$ 
at the time of the \Chandra observation.  The high source confusion
near the galactic center for the ASM makes this data relatively noisy,
so we take 
1.5 cts s$^{-1}$ as an upper limit on the X-ray flux.  Assuming the
same spectrum as we find in 2000 (see Sect. 2.5), 
we derive an unabsorbed 0.5-10 keV X-ray flux of $6.1\times10^{-10}$
ergs cm$^{-2}$ s$^{-1}$ and an unabsorbed J-band flux of $1.2\times10^{-13}$
ergs cm$^{-2}$ s$^{-1}$, thus log($L_X/L_J$)$\leq$3.7.  We note that
van Paradijs and McClintock (1995) find log$(L_X$(2-11
keV)/$L_{Opt}$(3000-7000 \AA))$\sim2.7$ for typical bright LMXBs. 

Rough $(J-K)_0$ colors have been reported for the
soft X-ray transients in outburst QX Nor, or X1608-52
($0.5<(J-K)_0<-0.4$, Wachter 1997), 
 and SAX J1808.4-3658
($0.1\pm0.1$, Wang et al.~2001) and the faint LMXB X1323-619
($-0.1\pm1.4$, Smale 1995).   Our value of (F110W-F187W)$_0\sim-0.17\pm0.3$
for star A, implying $(J-K)_0\sim-0.3$, is consistent with this broad
range.   Infrared colors observed by Wachter 
(1998) give an even broader range ($-0.4<(J-K)_0<1.41$), as some LMXBs
are dominated by the 
disk light in the IR, while others are dominated by the light of the
secondary. This makes identification of LMXBs by infrared colors very
difficult, and leaves our identification uncertain.   

\subsection{X-ray Classification and Completeness}

We extracted the numbers of counts within 1'' radius of the sources
(containing 90\% of source flux at 1.5 keV\footnote{CXC Proposer's
Observatory Guide, http://cxc.harvard.edu/udocs/docs/docs.html}) in the 
0.5-1.5, 0.5-4.5, and 1.5-6 keV bands.  We then subtract 1/3 (ratio of
areas) of the
counts from background 1''-2'' annuli around these sources for each
band (except where 
the annuli would overlap with other sources, where  
we offset the extraction region).  The results are listed in Table 1.
Constructing an X-ray CMD (as in GHE01a; Xcolor is defined as 2.5
log[0.5-1.5 keV counts/1.5-6 keV counts], while the vertical axis is
log[0.5-4.5 keV counts]), we are able to identify two 
populations of faint sources with distinct X-ray colors.  Using an $N_H$
of $1.2\times10^{22}$ cm$^{-2}$ (derived from CLG02's estimate of
A$_V$=6.75, using Predehl \& Schmitt's (1995) conversion factor), we
apply a uniform shift to the coordinates of the CMD as Pooley et
al. (2002b) do for the moderately reddened globular cluster NGC 6440.
The identified dim sources (and the color of the LMXB halo, which
determines the background) are plotted in the upper panel of Figure
5.   The upper and right axes  
show the raw (absorbed) Xcolor and counts, while the lower and left axes
represent the corrected (unabsorbed) Xcolor and counts.  (We note that the
interstellar medium decreases soft fluxes more than hard fluxes. Thus this
simple shift of axes should be used with caution.)  Another version of
the same diagram, but plotting luminosities derived from spectral fits
(section 2.5 below) against corrected Xcolor as above, is shown in the
lower panel of figure 5.  Errors on the vertical axis are derived
from errors in the counts, and do not include spectral
uncertainties. This version of the diagram (inspired by 
Becker et al.~2003), is more useful for comparing different clusters,
but is dependent on spectral and reddening assumptions.  
The luminosities of the two populations roughly agree with the 
range observed for their respective counterparts in 47 Tuc and NGC 6440.  This 
suggests that the soft population (W2, W3, W4, W8) should be
identified with qLMXBs containing neutron stars 
and that the harder population is composed of bright CVs.

We estimate our incompleteness, due to the X-ray halo and streak from the LMXB
EXO 1745-248, by the use of 
artificial-star tests.  We generated uniform grids of artificial point sources
at different luminosities using the MARX 
software\footnote{Available at
http://space.mit.edu/ASC/MARX.} and merged the simulated data with our
0.5-2.0 keV master image to see if we could detect the fake
sources. We scaled the average (Poisson-distributed) counts of the 
artificial sources to the average detected counts of real sources for 
which XSPEC fitting gave the target luminosity (see section 2.5
below).  E.g., for 
$L_X=10^{32}$ erg s$^{-1}$ sources, the average number of counts per
simulated source was 20, and for $L_X=3\times10^{32}$ erg s$^{-1}$ it
was 45 counts.  (The counts-luminosity relation varied depending on
the kind of sources prevalent at different luminosities.)  We performed the
tests for artificial source 0.5-2.5 keV luminosities of $10^{32}$,
$1.7\times10^{32}$, $3\times10^{32}$, and 
$10^{33}$ ergs s$^{-1}$, assuming spectra similar to the average of
our detected sources.  We separated the artificial sources 
by 5'' so that each source would suffer only the crowding that exists
in the real data.  We merged grids of point sources with the Chandra
image four times for the 
$10^{32}$ and $1.7\times10^{32}$ tests, and three times for the other
two tests, shifting the grids by several arcseconds between each
trial.  We then applied WAVDETECT to each image, keeping the same 
parameters as in section 2.2.  We counted the number of artificial sources that
landed within 
one optical core radius, between one and two core radii, and between two and
three core radii for each trial, and the number of sources
detected in each region, for a total of $\sim$1000 point-source
trials.  We note that this approach takes crowding into account, in
that the test sources will be crowded by any real undetected sources.
Parts of the 1-2 core radii annulus and 2-3 core radii 
annulus were imaged only in the 5 ksec exposure and not in the 42 ksec
exposure.  The 2-3$r_c$ bin also includes a small area that is completely
unexposed (due to the use of the subarray).  We scaled the MARX point
source luminosities appropriately 
and performed separate trials for the low-exposure regions (where
sources below 
$3\times10^{32}$ ergs s$^{-1}$ were never detected), and then included
these results in the total completeness calculations.  This allowed an
incompleteness factor to be determined for each radial annulus for each
of four luminosity bins, which we created by equally dividing
(in a logarithmic sense) the ranges between our test luminosities.

The number of real sources detected was too small to effectively
determine both a radial distribution and a luminosity function (1
source within 1 $r_c$, 5 within 1-2 $r_c$, 3 within 2-3 $r_c$).
Therefore we assumed a luminosity function of the form 
$dN \propto L^{-\gamma} d {\rm ln} L$, $L>L_{\rm min}$  following Johnston \&
Verbunt (1996).  We fixed $\gamma$ at 0.5
first, as found for sources between $10^{32}$ and $10^{33}$ ergs
s$^{-1}$ in NGC 6440 (Pooley et al.~2002b) and the general globular
cluster system (Johnston \& Verbunt 1996), and then allowed $\gamma$
to vary within the range 0.29 to 0.78 found by Pooley et al.~(2002b).
The uncertainty in our incompleteness due to different possible values of
$\gamma$ was small compared with errors due to small-number statistics.   
For each annulus, we determined an average incompleteness factor using
the results of our testing and the weights derived from the luminosity
function for the luminosity bins as specified above.

To calculate how many sources were missed, we take the number of
sources detected in each annulus and apply the 1$\sigma$ errors for
small numbers given by Gehrels (1986).  These numbers and errors are
multiplied by 1/f, where f is the completeness fraction, to obtain
the average density of sources in each annulus.  Then we calculated
the expected numbers of missed sources, using the estimated density
and again applying small-number errors.  We ignore possible 
systematics in the method, since the random errors are so large.  
Thus, above $10^{32}$ ergs s$^{-1}$, 
we find that within 1 $r_c$ we predict $0.8^{+2.8}_{-0.8}$ additional sources,
within 1-2 $r_c$ we predict $1.5^{+2.8}_{-1.3}$ additional sources, and within
2-3 $r_c$ we predict $1.1^{+2.6}_{-1.0}$ sources, giving a total
probable number of $11.4^{+4.7}_{-1.8}$ sources
between $10^{32}$ and $10^{33}$ ergs s$^{-1}$ in Terzan 5.  We note that
this source population is very similar to the 11 sources in this
flux range identified in NGC 6440 (Pooley et al.~2002b), as expected
considering the similar cluster parameters (as derived in section 3.2
below). 

The radial distribution we infer can be compared with that predicted
by a ``generalized King model'' (Lugger, Cohn, \& Grindlay 1996) where
the surface density of sources varies as 

$S_x(r)\propto[1+(\frac{r}{r_c})^2]^{\frac{\alpha_X}{2}}$

\noindent
where $\alpha_X=-(3q-1)$ and $q=\frac{<m_x>}{<m_*>}$, the ratio of the
average X-ray system mass to the mass of the stars that dominate the
density profile (i.e., turnoff-mass stars for King model clusters).  The soft
sources in 47 Tuc (thought to be largely MSPs) have $\alpha_X=-3.6$ and the CVs
in 47 Tuc have $\alpha_X=-3.1$ (Grindlay et al.~2002).  Comparing the corrected
source counts in each annulus with the predictions of generalized King
models with $\alpha_X=-2$, $-3$, and $-4$, we find a best fit with
$\alpha_X=-2$, suggesting 
similarity in source mass to the turnoff mass stars.  However,
the large uncertainties on our source numbers (especially in the core)
allow consistency with $\alpha_X=-3$ in a $\chi^2$
sense.  We conclude 
that the small number of sources does not allow a sensitive test of
the radial distribution, and thus of the mass of the X-ray sources.
Upcoming \Chandra observations of Terzan 5 (PI: R. Wijnands) should
identify more X-ray sources, allowing a better probe of the luminosity
function and radial distribution.

\subsection{Spectral fitting and variability of faint sources}

We extracted the spectra of the detected X-ray sources (except for EXO
1745-248; see section 2.6) from 1'' radius
circles around the wavdetect positions, with background spectra taken
from 1-2'' annuli, using the {\it psextract} CIAO script.  We utilize
the recently released ancillary response file correction\footnote{See
http://cxc.harvard.edu/cal/Links/Acis/acis/Cal\_prods/qeDeg/index.html.}
for the time-variable low-energy quantum efficiency degradation in all
spectral fits in this paper.  This correction had a very small impact
on our fits (in part because of the large $N_H$).   We group the
spectra with at least 10 counts/bin, and fit models in XSPEC
(Arnaud 1996), all using 
the near-infrared-derived value of photoelectric absorption
(CLG02). The models used were thermal
bremsstrahlung, a powerlaw, and a hydrogen atmosphere appropriate for
an accreting neutron star in quiescence (Lloyd 2003).  
Simple background subtraction is not the preferred way to fit
spectra from sources with high background, due to deviations from
Poisson statistics in the subtracted counts.    We
verified that Sherpa spectral fitting of source and background
simultaneously gives results within the errors of our XSPEC analysis where
we are able to use the same models, but
note that these low signal-to-noise spectra should be taken with caution.

A thermal bremsstrahlung spectrum of kT$\gsim$5
keV seems to describe most cataclysmic variables in the globular
clusters 47 Tuc and NGC 6397 (Grindlay et al.~2001a,b). We fixed
the normalization of the hydrogen atmosphere models, 
set to (r [km]/D[10 kpc])$^2$= 132 (appropriate for a 10 km neutron
star at 8.7 kpc), and the redshift to 0.306.  This tests for
compatibility with a canonical 1.4 \Msun,
10 km neutron star.   A hard power law component has been found to
dominate qLMXB spectra above 2-3 keV in some systems, but has not yet
been identified in globular cluster qLMXBs that have not displayed
outbursts (see Heinke et al.~2002).  We make no attempt to confirm or rule
out such a hard component in this paper due to the extremely high
background (from EXO 1745-248) above 2 keV.    

The results of our spectral fitting are shown in Table 2 (errors are
90\% confidence).  All of
the four sources on the right side of the X-ray CMD (W2, W3, W4, W8)
are well fit by our hydrogen atmosphere model.  Alternative models
require steep powerlaw spectra 
($\alpha>$2) or bremsstrahlung temperatures below 2 keV, not seen in
bright ($>10^{32}$ ergs s$^{-1}$) CV
 or millisecond pulsar systems (Grindlay et al.~2001a, 2001b, 2002, Pooley et
al. 2002a).  Their high luminosities of 
$3-6\times10^{32}$ ergs s$^{-1}$ and soft spectra rule out flaring
coronal events from BY Dra or RS CVn 
 binary systems (see e.g. Grindlay et al.~2001a).   Therefore we
conclude that these are probably qLMXB systems.  The five sources on
the left of the CMD (W5, W6, W7, W9, W10) are well fit by
bremsstrahlung spectra with kT$>$5 
keV, or power law spectra with $\alpha$=1.0-1.6, and are not acceptably
fit by hydrogen atmosphere models with the normalization fixed to a 10
km radius.  We conclude that these are most likely highly X-ray
luminous cataclysmic variables, though MSPs and RS CVn stars are
(less likely) possibilities.    
 
We test for variability by extracting the events files between 0.5 and
2.5 keV within 1'' of each source, and using the IRAF PROS task
\texttt{vartst}.  This task applies the Kolmogorov-Smirnov and Cramer-Von
Mises tests (Daniel 1990) to attempt to disprove the hypothesis that the source
flux is constant. 
We choose only events below 2.5 keV because the events at higher
energies are generally dominated by the background from EXO 1745-248.  Only two
sources are found to be significantly variable in both tests, W9 and
W10 (identified as probable CVs above).  Both are
variable with 99\% confidence.  This is consistent with the
trend for qLMXBs of $L_X\sim10^{32-33}$ ergs s$^{-1}$ to show generally
constant flux, compared to globular cluster CVs that are often highly
variable (Grindlay et al.~2001b, Heinke et al.~2002, but cf. Becker et
al. 2003). 

\subsection{Spectral fitting of EXO 1745-248}

  When observed by the ROSAT All-Sky Survey at a luminosity of
log($L_X$)=35.3, the 0.5-2.5 keV spectrum of EXO 1745-248 was well fit
by a powerlaw of photon index $\alpha$=1.2, using a 
hydrogen column of $N_H=2\times10^{22}$ cm$^{-2}$ (Verbunt et
al. 1995).  A ROSAT HRI observation in March 1991 found it at a luminosity
of log($L_X$)=34.6, assuming the same spectrum (Johnston et al.~1995).  
Only a broad-band X-ray spectrum can provide detailed information
about the nature of the LMXB, by simultaneously constraining the
photoelectric absorption column, any iron lines or edges, and the
overall spectral shape.  Fortuitously, there exist \RXTE data
simultaneous with our \Chandra data, which we also analyze.

To study the \Chandra spectrum of EXO 1745-248 in outburst, we first 
selected an annulus from 5  
to 9 arcseconds from our best-determined position, and removed the
portion affected by the readout streak.  This annulus is not affected
by pileup, and includes only one faint source (W4) which should not
significantly affect the spectral fitting (contributing $\sim17$ of
24739 counts).  
We extracted the spectrum using bins of width 75 eV (oversampling the
energy resolution) and using a background region on the same chip 2'
away.  We also extract a lightcurve from the same region.  We see a
gradual rise from 0.4 counts s$^{-1}$ to 0.5 counts 
s$^{-1}$ during the first observation, while the second (5 ksec) observation
five days later shows a rate of 0.9 counts s$^{-1}$ (consistent with
the rise in the ASM countrate).  Flickering is clearly seen, but we
put off detailed study of temporal variability for future work.
An absorbed powerlaw of photon index 0.24 provides a good fit to the
continuum, with 
a strong broad iron line (EW 440$^{+190}_{-230}$ eV, $\sigma=0.5\pm0.1$ keV)
required at 6.7 keV.  However, we note that the spectrum will be
altered both by the effects of dust scattering and by the energy
dependence of the \Chandra mirror psf (e.g., Smith,
Edgar \& Shafer 2002).  Therefore we extracted the spectrum from the readout 
streak, which does not suffer this hardening since the counts are
recorded in the image core.  

We used the CIAO task {\it acisreadcorr} to identify and reposition
events in the 
EXO 1745-248 readout streak. We use a strip of width 2 pixels and
omit a region of 25 pixels radius around the source location, to avoid
photons from the X-ray halo and regions that suffered pileup.
Pileup is not an issue in the selected region, as the 8750 counts
selected give 0.15 cts  
(0.841 s CCD frame)$^{-1}$, spread along a 78$\times$2 pixel column.  We
extracted those events using the CIAO script {\it psextract}, and chose a
background region adjacent to the strip (adjusting the
BACKSCAL parameter by hand).  We
 ignore data above 10 keV and below 0.7 keV (which is almost
entirely background photons, considering the high absorption to this
source).  We fit this \Chandra readout streak spectrum
along with the simultaneous \RXTE data to understand the full spectral shape.

From the \RXTE data, only the PCA STANDARD 2 data and the HEXTE Archive
data (from both 
HEXTE clusters separately) are analyzed in this paper.  We choose
time intervals when 
three of the PCA units were on (the maximum during the observation)
and the elevation above the earth's limb was greater than 10 
degrees.  Selecting only the
top xenon layer data from the PCA, we use 
PCABACKEST version 3.0 (released Feb. 1, 2002) and PCARSP version
8.0, and correct the PCA data manually for deadtime.      
 We used the HEXTE response matrices hexte\_97mar20c\_pwa.rmf
and hexte\_97mar20c\_pwb013.rmf, and
corrected the HEXTE data for background and deadtime using the
HXTBACK and HXTDEAD (version 2.0.0) scripts.  We eliminate 200 seconds
around a possible Type I X-ray 
burst, which we do not investigate in this paper (Type I bursts are
common from EXO 1745-248; see Inoue et al.~1984).  This gave total
corrected exposure times of 2060, 731, and 777 seconds for the PCA and
HEXTE clusters A and B respectively.  
We analyze PCA data from 3 to 25 keV, and HEXTE data from 25 to 125 keV.
Following Barret et al. (2001), we add systematic errors of 0.5\% to 
the PCA data below 15 keV, and 1\% to PCA data above 15 keV using the
FTOOLS GRPPHA.  We leave
an overall normalization free between the \Chandra, PCA and HEXTE data, but
link all other parameters between them in a joint fit.  (The PCA and HEXTE
relative normalizations are generally not well-calibrated; we find
the HEXTE normalization to be 40\% lower than the PCA normalization.)

The standard models for fitting neutron star LMXB spectra are an
absorbed multicolor blackbody, or a simple blackbody, with an additional
hard component due to comptonization of soft photons by hot electrons
(generally 
assuming, as here, a spherical geometry); see Barret et al.~(2000),
and Sidoli et al.~(2001).  Fitting a simpler model to the RXTE data,
consisting of DISKBB plus a 
gaussian and powerlaw, gave a photon index of 1.5, but failed to fit
the HEXTE data due to a dropoff of flux at high energies
($\chi^2_{\nu}$=8.1 for 230 degrees of freedom).
 We begin by fitting our \RXTE spectra with a model consisting of an absorbed
multicolor blackbody 
(DISKBB; Makishima et al.~2000), a comptonization model 
(COMPTT; Titarchuk 1994), and the Fe-line gaussian, and with another model
consisting of a blackbody, COMPTT, and the gaussian.  The \RXTE fits for
both models are
significantly improved by the addition of a smeared iron edge
(``smedge''; Ebisawa et al.~1994) near 8 keV, as expected for
reflection of Comptonized hard X-rays from a disk. 
(An F-test gives a probability of $10^{-4}$ that the smeared edge is not
needed.)    For the rest of this analysis, the ``standard model'' shall 
refer to PHABS(DISKBB + COMPTT + GAUSSIAN)*SMEDGE*CONSTANT.  The data/model
ratio for PCA data, fit with the standard 
model with the normalizations of the gaussian and smeared edge set to
zero, are shown in Figure 6. 

The \Chandra spectrum extracted displays
what appear to be emission lines at $\sim$1.95 and 2.1 keV.  These
features are not apparent in either 
the adjacent background, or the annulus spectrum described above.  
Similar features are also seen in the readout streak spectrum of GX
13+1 (Smith et al.~2002), and in the
high count rate continuous-clocking spectrum of RX J170930.2-263927
(Jonker et al.~2003).  The best-fitting standard model to the \RXTE
data does not give an acceptable fit to the \Chandra data above 5 keV,
where they overlap, nor does any other model fit both the \Chandra and
\RXTE data; see Figure 7. For the best-fit standard model, the reduced 
$\chi^2$=1.41 for 225 degrees of freedom (null hypothesis
prob.=$5.7\times10^{-5}$).  Note that the \Chandra spectrum 
appears to have been shifted in energy compared to the model
predictions.  We understand this effect as being due to a gain shift
between the calibrated response of the timed exposure mode, and the actual
response to events  
occurring during the readout period, when the voltages in the CCD are
expected to be different.  By measuring the difference between the
model and data edges near 2 keV, and the model (derived from \RXTE
fits) and data between 5 and 9 keV, we estimate the gain shift at
7$\pm1$\%.  

We attempt to compensate for this gain shift by increasing the size of
the energy bins in the events file by 7\%, to 15.6 eV from 14.6
eV, while keeping the same response, ancillary response, and background
files and extracting the altered-gain spectrum in the same way.  Fitting
this shifted-energy spectrum, along with the \RXTE spectra, gives a
greatly improved fit (see Figure 8).  The standard model fit to these
spectra gives a reduced $\chi^2$=1.06 for 225 dof (prob=0.250).  The
derived parameter ranges are robust; spectral fits to \RXTE data alone
give similar parameter values (with larger uncertainties for the soft
components, and $N_H$ virtually unconstrained).
Although this simple calibration must be used with a great amount of caution,
we believe that the quality of the resulting spectral fits supports
our decision to utilize it for this analysis.  

We list the 
best-fitting parameters (and 90\% confidence errors) for the two
models in Tables 3 and 4, along with 
the derived values for the radii of the thermal components and the ratio
of luminosities in the two components {\it f}.   We have not corrected
the multicolor disk-blackbody spectral parameters for the spectral
hardening factor expected in high accretion rate systems (see Shimura
\& Takahara 1995, Merloni et al.~2000), for ease of comparison to the 
work of SPO01 and Parmar et al.~(2001) below.  The averages of the
best-fit luminosities (from the two  
fits' PCA normalizations) are $L_X$(0.5-10)=$2.1\times10^{37}$, 
and $L_X$(0.1-100)=$6.6\times10^{37}$ ergs s$^{-1}$.  Several authors
(e.g. Bloser et al.~2000) have commented upon an absolute uncertainty
in the PCA flux normalizations of order 15\%, which we do not attempt
to compensate for. 
We note that the ranges of $N_H$ required by these fits
(1.3--$1.9\times10^{22}$ cm$^{-2}$) are slightly larger than the 
infrared-derived estimate of CLG02, $1.2\times10^{22}$, indicating
probable internal absorption in the LMXB system.  

\section{Discussion}
\subsection{Ultra-compact nature of EXO 1745-248}

We compare our results from fitting combined \Chandra and \RXTE data with
our ``standard model'' spectrum, to the spectral fits of SPO01 and
Parmar et al.~(2001) for the other globular cluster LMXBs,  
observed with BeppoSAX.    Five comparisons using the DISKBB + COMPTT model 
in SPO01 separate the ``normal'' (not ultra-compact) globular cluster LMXBs
in Terzan 2, NGC 6440, NGC 6441, and Terzan 6 from the ultra-compact
(binary periods $<$ 1 hour)
LMXBs in NGC 6624, NGC 1851, NGC 6712, and (probably) NGC 6652.  SPO01
note that these comparisons suggest that the DISKBB model is 
physically meaningful only for the ultracompact LMXBs.

$\bullet$ The ultracompact LMXBs show
an inner DISKBB temperature below 1 keV, while the normal LMXBs show
kT$_{\rm in}$ between 1.9 and 3.5 keV. EXO 1745-248 shows kT$_{\rm
in}=0.80^{+.38}_{-.16}$ keV. 

$\bullet$ The seed photon temperature kT$_0$
for the ultracompact binaries is roughly equal to the inner DISKBB
edge temperature, while for the normal LMXBs the seed photon
temperature is 4--5$\times$ lower. EXO 1745-248 seems to have similar
values for kT$_0$ (1.29$^{+.40}_{-.13}$ keV) and kT$_{\rm in}$
(0.80$^{+.38}_{-.16}$ keV).   

$\bullet$ The inferred inner radius 
R$_{\rm in}$ (cos i)$^{0.5}$ of the normal LMXBs is very small (0.3 to
1 km), while for the ultracompact LMXBs it is 3 to 45 km.  EXO
1745-248's implied R$_{\rm in}$ (cos i)$^{0.5}$ is 
8.7$^{+5.0}_{-4.4}$ km.  

$\bullet$ The seed photon
emission radius $R_W=3\times10^4 d \sqrt{f_{comptt}/(1+y)}/(kT_0)^2$
km, where $d$ is the source distance in kpc, $f_{comptt}$ is the
luminosity in the Comptonized component in ergs cm$^{-2}$ s$^{-1}$,
and $y$ is the COMPTT 
Comptonization parameter, $y=4 kT_e \tau^2/m_e c^2$. $R_W$ is 
similar to the inner disk radius R$_{\rm in}$ (cos i)$^{0.5}$ in
the ultracompact LMXBs, while it is $\sim$10-50$\times$ larger for the
normal LMXBs.  For EXO 1745-248 we calculate $R_W$=5.4$^{+1.3}_{-2.9}$
km, similar to its R$_{\rm in}$ (cos
i)$^{0.5}=8.7^{+5.0}_{-4.4}$ km.  (We also note that $y=5.2^{+0.5}_{-0.6}$ 
for EXO 1745-248, within the range 1-7 of most of the LMXBs studied in
SPO01).

$\bullet$ Finally, the
implied mass of the compact object, derived from the DISKBB
extrapolated bolometric luminosity and the inner disk temperature
(following  Makishima et al.~2000, including spectral hardening), is
consistent with 1-3 \Msun\ for the 
ultracompact binaries, but is nearer 0.1 \Msun\ for the normal LMXBs.
We derive 1.6$^{+0.8}_{-0.9}$ \Msun for EXO 1745-248.

Each of these comparisons strongly indicates that EXO 1745-248 is an  
ultracompact LMXB.   Some of these spectral
differences have been used by Parmar et al. (2001) to identify the
bright LMXB in NGC 6652 as a probable ultracompact system, which is
supported by the short optical periods proposed by Heinke et al.~(2001).
The identification of EXO 1745-248 as a probable ultracompact LMXB
 makes five of 13 luminous LMXBs in globular clusters ultracompact, a far
greater proportion than in the field.  This 
strengthens the conclusion of Deutsch et al.~(2000) that dynamical
effects are probably responsible for the generation of ultracompact
LMXBs in globular clusters.

Another interesting result is the strength of the ionized iron line in
EXO 1745-248, with an equivalent width (EW) $\sim190$ eV.   This is many   
times larger than the average upper limit (25.6 eV) on the equivalent width of
an iron line for the nine globular cluster LMXBs with good statistics
reported by SPO01 and Parmar et al.~(2001).  Only two other globular cluster  
LMXBs have identified iron lines; 4U 1820-30 in NGC 6624
(EW=13$^{+12}_{-11}$ eV, SPO01; EW=31$^{+12}_{-11}$ eV at 6.6 keV,
Asai et al.~2000; EW=27-94 eV in numerous \RXTE observations, Bloser et
al.~2000) 
and Terzan 2 (EW 21$^{+10}_{-9}$ or 27$\pm10$ eV depending on
continuum model, Barret et al.~2000). Among these LMXBs, only 4U
1820-30 in NGC 6624 has a reported reflection edge feature (at 7.7 or
8.9 keV; Bloser et al.~2000), in its low or ``island'' state. The
strength of the iron line may be increased by the high  
metallicity of Terzan 5, suggested by CLG02 to be greater than solar,
compared to the significantly subsolar metallicity for the other
globulars with LMXBs except Liller 1.   We note that the
6.55$^{+.06}_{-.07}$ keV energy and $\sigma=0.31^{+.16}_{-.15}$ keV
breadth of this line are similar to those of field neutron star  
LMXBs reported in Asai et al.~(2000), 6.56 keV and 0.2$\pm0.1$ keV
(0.5 keV FWHM) respectively, although this line is somewhat stronger
than in most field LMXBs.  The iron lines
studied by Asai et al.~(2000) were interpreted as radiative
recombination in an accretion disk corona, with the breadth being due
primarily to Compton scattering, with possible small contributions from
Doppler shifts (due to Keplerian motion near the NS) and line mixing
from plasmas in different ionization states. The location
of our line and smeared edge suggest iron that is less highly ionized
than He-like Fe XXV K lines (6.68 keV line and 8.8 keV edge).  Disk
reflection from carbon-like Fe XXI (6.54 and 8.3 keV line and edge) is
more consistent with our line and edge energies.  Further analysis of the 
\RXTE outburst spectrum and variability of EXO 1745-248 is being performed
by Homan et al.~(in preparation).
   
Kuulkers et al. (2002) have recently presented a BeppoSAX spectrum of
EXO 1745-248, taken 15 days later during the same 2000 outburst.
Kuulkers et al. (2002) find an interstellar 
absorption of $N_H=3.8^{+0.9}_{-0.7}\times10^{22}$ cm$^{-2}$ when using the
DISKBB + COMPTT model, and $N_H=2.3^{+0.6}_{-0.8}\times10^{22}$ using
the BBODYRAD +COMPTT model.  These values are inconsistent with our
(more precise) values for $N_H$, suggesting that the absorption column
may have changed during the outburst. Their $N_H$ values are
inconsistent with the 
 optically derived extinction value (CLG02; see figure 1 in Kuulkers
et al.~2002), requiring an absorption column within the LMXB system
ten times larger than
shown in any other globular cluster LMXB.  If this were the case, we
would expect a high inclination and significant dips due to variable
absorption. While these are not seen in our Chandra and
single-observation \RXTE lightcurves, R. Wijnands notes (priv. comm.)
that such dips are indeed seen in the full \RXTE lightcurves of EXO
1745-248's 2000 outburst (J. Homan, in prep.).  This may imply that the
system is at high inclination, though the observed disk reflection
component suggests a low inclination.  We also note that their spectral
parameters, while not agreeing with ours in every detail, support our claim
that EXO 1745-248 is ultracompact, particularly in their small value
of kT$_{in}$ for their DISKBB + COMPTT fit. 

\subsection{Terzan 5 cluster parameters}

It has long been suspected  that Terzan 5 has one
of the highest rates of close encounters between stars of any Galactic
globular cluster.  The ``collision rate'', or rate of close
encounters given by 
$\Gamma \propto \rho_0^2 r_c^3 / \sigma$ (where $\rho_0$ is the
central density, $r_c$ is the core radius, and $\sigma$ is the
central velocity dispersion), is expected to predict
the relative rates of formation of accreting binary neutron star
systems by two-body encounters (Verbunt \& Hut 1987).  Thus, the
similar numbers of X-ray sources in NGC 6440 and Terzan 5 might seem a
surprise, as Terzan 5 has been predicted to show three times as many
collision products as NGC 6440 and 17\% of the total of such objects in the
Galactic globular system (Verbunt 2002).

Our calculation of the central density of Terzan 5 uses the
extinction-corrected central surface brightness 
$\mu_V(0)=20.5$, combining the star-count profile of CLG02 for the
inner core with the
surface brightness profile of Trager et al. (1995) for normalization
beyond 10''.  We use the central 
concentration parameter c=2.0, core radius $r_c$=7\farcs9,
heliocentric distance of 
8.7 kpc, and A$_V$=6.75 from CLG02, as well as M$_V=-7.91$ from the
updated Harris (1996; rev. 1999) catalog.  Following the prescription
of Djorgovski (1993), our result is that  
Terzan 5's central density is $1.7\times10^5 L_{\odot}$ pc$^{-3}$,
significantly less than given by Djorgovski
(1993), Harris (1996), or CLG02.  The former two studies used a larger
value of $A_V$, which produces a larger correction of the surface
brightness.  CLG02 had scaled the central surface brightness value of
Djorgovski (1993) for changes in the profile, without updating the
value of $A_V$.  We note
that this calculation is consistent with the lower limit of $5.0\times10^5$
\Msun pc$^{-3}$ derived by Lyne et al. (2000) from the
acceleration-induced \.{E} of
pulsar Terzan 5 C, provided that M/L$\geq 3.0$.  Using this
central density, and the distance and core radius from CLG02, the
collision rate 
in Terzan 5 is similar to that in NGC 6440 (5.9\% of the
total collision rate in the globular system),
instead of three times larger.  This is consistent with the results of
our artificial star tests (section 2.3), which suggest that Terzan 5 may
contain $\sim$11 sources in the range $10^{32-33}$ ergs s$^{-1}$, compared to
11 in NGC 6440. A full analysis of the luminosity
function and density 
weighting of quiescent LMXBs in globular clusters will be presented in
Heinke et al.~(2003, in prep.)  We also revise the estimate of the
central relaxation time of Terzan 5 to $2\times10^8$ years from
CLG02's value of $4\times10^7$ years, following Djorgovski's (1993)
scaling of $t_{rc}\propto\rho_0^{0.5}r_c^3$. 
This indicates that the cluster is not as close to the verge of core
collapse as suggested by CLG02.  The high density of compact binaries
in the core 
seems to be due primarily to the high density of the cluster's massive
core.  The large numbers of millisecond pulsars (60-200) in the core
of Terzan 5 estimated by Fruchter and Goss (2000) would require a long
period of high core density to form the millisecond pulsar
progenitors.  MSPs may be most efficiently formed early in the
cluster history by intermediate-mass X-ray binaries
(e.g. Davies \& Hansen 1998). Therefore it seems likely that Terzan 5's MSPs
were formed early in its history, as were those in 47 Tuc (Grindlay et
al. 2002), which has a roughly similar inferred ratio ($\sim15-50$) of
MSPs to qLMXBs. 

\section{Conclusions}

We have presented a reasonable infrared candidate to EXO 1745-248 in Terzan 5,
identified by its blue color and positional coincidence with the
boresighted \Chandra position.  We have assembled a broad X-ray
spectrum using a simultaneous \RXTE observation and the \Chandra spectrum from
the readout streak, slightly altering the energy scale of the \Chandra
readout spectrum to account for observed gain variation during the
readout. We utilized the 
empirical comparisons of SPO01 to indicate that this LMXB appears 
similar to other ultracompact LMXBs in globular clusters, suggesting
that EXO 1745-248 is the fifth ultracompact LMXB known in a globular cluster.
We also identify a 
broad, strong 6.55 keV iron line, the strongest
(EW=188$^{+86}_{-83}$ eV) yet discovered in a
globular cluster LMXB, with an accompanying smeared $\sim8.1$ keV iron
edge.

The superb resolution of \Chandra has allowed us to identify nine
faint X-ray sources within 30'' of an LMXB in outburst in Terzan 5.
Spectral fitting with bremsstrahlung and powerlaw models, and a
 neutron star hydrogen atmosphere model (Lloyd 2003), suggests that
four of these sources are qLMXBs, while five are candidate CVs. 
Artificial point source testing suggests that we are missing
$\sim$30\% of the sources in the range $L_X(0.5-2.5\ {\rm
keV})=10^{32-33}$ ergs s$^{-1}$ due to the 
presence of the LMXB in outburst. This implies a total cluster population of
$11.4^{+4.7}_{-1.8} (1\sigma)$ sources with $L_X>10^{32}$ ergs s$^{-1}$
(excluding the LMXB).   A recalculation 
of the central density of Terzan 5 from updated cluster parameters
gives log($\rho_0$)=5.23, suggesting that Terzan 5 is not the
richest of the globular clusters in stellar encounter products and is
not as dynamically unstable as previously thought (CLG02).  Thus we find that
the numbers of X-ray sources in Terzan 5 are consistent with the
numbers discovered in other globular clusters and the currently
favored formation methods.

Upcoming \Chandra observations of Terzan 5 and NGC 6440 (PI:
R. Wijnands) will allow us
to better constrain the variability and spectra of the qLMXBs in those
clusters.   $V$ and $I$ observations of Terzan 5, with
the \HST Advanced Camera for Surveys in HRC mode at times when
EXO 1745-248 is in outburst vs. quiescence, could 
unambiguously verify the identification proposed here.

\acknowledgments

C. H. thanks J.~M. Miller, R. Wijnands, T. Gaetz, R.~K. Smith,
R. Edgar, and the anonymous referee 
for useful comments that have improved this paper.  CH also thanks
S. Wachter for access to unpublished data.  
This work was supported in part by \Chandra grant GO0-1098A and \HST
grant GO-7889.01-96A.  \RXTE data and results 
provided by the ASM/\RXTE teams at MIT and at the \RXTE SOF and GOF at NASA's
GSFC. The Guide Star Catalogue-II is a joint project of the Space
 Telescope Science Institute and the  Osservatorio Astronomico di
   Torino.  This research has made use of the data and resources
obtained through the HEASARC on-line service, provided by NASA-GSFC,
the VizieR catalogue access tool, CDS, Strasbourg, France, 
and NASA's Astrophysics Data System.

\hspace{-1cm}
\begin{deluxetable}{lccccr}
\tablewidth{7.2truein}
\tablecaption{\textbf{Names, Positions and Counts of Detected Sources}}
\tablehead{
\colhead{\textbf{Source Name (Label)}}  &
\colhead{RA} & \colhead{Dec} & \colhead{0.5-4.5 keV} &
\colhead{0.5-1.5 keV} &
\colhead{1.5-6 keV} \\
 & (17:) & (-24:) & (counts) & (counts) & (counts)
}
\startdata

EXO 1745-248 (LMXB)  & 48:05.196$\pm$.015 & 46:47.40$\pm$.20 & - & - & - \\
CXOGLB J174806.1-244642.9 (W2)  & 48:06.154$\pm$.005 & 46:42.68$\pm$.07 & 37$\pm$10 & 11$\pm$4 & 20$\pm$10 \\
CXOGLB J174805.3-244637.7 (W3)  & 48:05.370$\pm$.004 & 46:37.66$\pm$.05 & 77$\pm$13 & 30$\pm$6 & 41$\pm$13 \\
CXOGLB J174804.7-244644.6 (W4)  & 48:04.799$\pm$.003 & 46:44.64$\pm$.06 & 17$\pm$19 & 15$\pm$7 & -1$\pm$21 \\
CXOGLB J174804.4-244638.2 (W5)  & 48:04.402$\pm$.006 & 46:38.15$\pm$.09 & 81$\pm$11 & 13$\pm$4 & 80$\pm$12 \\
CXOGLB J174804.3-244703.8 (W6)  & 48:04.366$\pm$.004 & 47:03.75$\pm$.07 & 180$\pm$16 & 22$\pm$5 & 208$\pm$17 \\
CXOGLB J174804.2-244641.8 (W7)  & 48:04.214$\pm$.005 & 46:41.79$\pm$.09 & 85$\pm$11 & 10$\pm$4 & 86$\pm$12 \\
CXOGLB J174804.2-244648.4 (W8)  & 48:04.225$\pm$.05 &  46:48.34$\pm$.2 & 34$\pm$12 & 15$\pm$5 & 30$\pm$12 \\
CXOGLB J174804.0-244640.5 (W9)  & 48:04.059$\pm$.009 & 46:40.53$\pm$.06 & 50$\pm$10 & 4$\pm$3 & 53$\pm$11 \\
CXOGLB J174803.5-244649.2 (W10) & 48:03.539$\pm$.005 & 46:49.24$\pm$.08 & 69$\pm$11 & 8$\pm$4 & 75$\pm$12 \\

\hline
\multicolumn{6}{c}{Sources not associated with Terzan 5} \\ 
\hline
CXOU J174803.3-244854.1 (Star 1)  & 48:03.334$\pm$.004  &
48:54.04$\pm$.06  & 45$\pm$8 & 36$\pm$7  & 9$\pm$5  \\
CXOU J174751.7-244657.4  & 47:51.726$\pm$.010  &
46:57.42$\pm$.11  & 36$\pm$7 & 9$\pm$4 & 35$\pm$7  \\  
CXOU J174814.7-244802.2  & 48:14.703$\pm$.011  &
48:02.21$\pm$.12  & 29$\pm$7 & 7$\pm$4 & 28$\pm$6  \\ 
CXOU J174812.6-244811.1 (Star 2) & 48:12.649$\pm$.010  &
48:11.10$\pm$.13  & 28$\pm$8 & 20$\pm$6 & 6$\pm$5  \\

\enddata
\tablecomments{Sources detected in and around Terzan 5.  RA and
Declination values for cluster members begin with 17: and -24:,
respectively.  Errors are centroiding errors within the (shifted to
match GSC 2.2) 
\Chandra frame, and do not include systematic errors incurred in
matching to other frames (absolute errors perhaps 0\farcs5; see text).
Counts in each energy band were determined by subtracting the average 
count rate in surrounding 2'' annulus from the counts in 1'' circle
around source position (leading to one negative entry).  
}
\end{deluxetable}

\begin{deluxetable}{cccccccc}
\tablewidth{5.5truein}
\tablecaption{\textbf{Spectral Fits to Faint Sources}}
\tablehead{
\colhead{\textbf{Source}} & \multicolumn{2}{c}{H-atmosphere}  &
\multicolumn{2}{c}{Bremsstrahlung} & 
 \multicolumn{2}{c}{Powerlaw} & \colhead{0.5-2.5 $L_X$} \\
 & (kT, eV) & ($\chi^2_{\nu}$/dof) & (kT, keV) & ($\chi^2_{\nu}$/dof)
& ($\alpha$) & ($\chi^2_{\nu}$/dof) & (ergs  s$^{-1}$)
}
\startdata
W2 & 98$^{+5}_{-6}$ & 1.19/14 & 0.7$^{+0.5}_{-0.3}$  & 1.16/13 & 3.9$^{+1.1}_{-1.0}$ & 1.31/13 & 4.1$\times10^{32}$  \\
W3 & 104$^{+5}_{-5}$ & 1.36/22 & 1.0$^{+0.7}_{-0.3}$ & 1.05/21 & 3.1$^{+0.8}_{-0.6}$ & 1.20/21 &  5.6$\times10^{32}$ \\
W4 & 90$^{+9}_{-17}$ & 1.37/11 & 0.5$^{+2.5}_{-0.5}$ & 1.49/10 & 4.6$^{+\infty}_{-2.6}$ & 1.5/10 &  2.7$\times10^{32}$ \\
W5 & 92 & 3.0/16 & 6.3$^{+71}_{-3.8}$ & 1.29/15 & 1.7$^{+0.5}_{-0.5}$
& 1.34/15  & 1.9$\times10^{32}$  \\
W6 & 100 & 4.42/33 & $>29$ & 1.08/32 & 1.0$^{+0.3}_{-0.2}$ & 1.05/32 &  3.5$\times10^{32}$ \\
W7 & 90 & 2.50/16 & $>6$ & 0.51/15 & 1.2$^{+0.5}_{-0.5}$ & 0.51/15 &  1.5$\times10^{32}$ \\
W8 & 92$^{+6}_{-9}$ & 0.84/19 & 0.8$^{+1.7}_{-0.4}$ & 0.78/18 &
3.3$^{+1.8}_{-1.3}$ & 0.82/18 &  3.1$\times10^{32}$ \\
W9 & 79 & 2.27/13 & $>2$ & 1.37/12 &
1.3$^{+0.9}_{-0.8}$ & 1.37/12  &  8.6$\times10^{31}$ \\
W10 & 88 & 2.92/15 & $>7$ & 0.95/14  & 1.1$^{+0.6}_{-0.6}$ & 0.94/14 &  1.3$\times10^{32}$ \\
\enddata
\tablecomments{Spectral fits to faint cluster sources, with background
subtraction, in XSPEC.  All fits include photoelectric absorption
fixed at the cluster $N_H$, 1.2$\times10^{22}$ cm$^{-2}$.  Hydrogen
atmosphere fits are made with radius fixed to 10 km.  X-ray
luminosities are unabsorbed for the range 0.5 to 2.5 keV, from
hydrogen-atmosphere NS fits (W2, W3, W4, W8) or thermal bremsstrahlung fits. 
Errors in all the tables are 90\% confidence for a single parameter.}
\end{deluxetable}

\begin{deluxetable}{lcccccccr}
\tablewidth{6.0truein}
\tablecaption{\textbf{Spectral Fits to LMXB EXO 1745-248 Continuum}}
\tablehead{
\colhead{\textbf{Model}} & \colhead{$N_H$}   &
\multicolumn{3}{c}{COMPTT} & 
 \multicolumn{3}{c}{BB or DiskBB\tablenotemark{a}\ } & \colhead{$\chi^2_{\nu}$/dof} \\
 & (10$^{22}$ cm$^{-2}$)  & (kT$_0$) & (kT$_e$) &
($\tau_p$ ) & (kT) & (R, km) &
(f\tablenotemark{b}\ ) & \\
}
\startdata
DBB  & $1.72^{+.20}_{-.17}$  & 1.29$^{+.40}_{-.13}$ & 10.2$^{+.5}_{-.4}$ &
8.1$^{+.3}_{-.4}$ & $0.80^{+.38}_{-.16}$ & $8.7^{+5.0}_{-4.4}$ & 0.12 &
1.06/225 \\
BB & $1.46^{+.19}_{-.17}$  & 1.21$^{+.24}_{-.09}$ & 10.2$^{+.5}_{-.4}$ &
$8.1^{+.3}_{-.2}$ & 0.53$^{+.05}_{-.06}$ & 20.6$^{+6.2}_{-5.1}$ & 0.07  & 1.06/225 \\

\enddata
\tablecomments{Spectral fits to the LMXB EXO 1745-248 from \RXTE data and the
\Chandra readout streak with energy scale alteration (see text), with 
background subtraction in XSPEC.  A gaussian, a smeared edge, 
photoelectric absorption, and comptonization 
using the XSPEC model COMPTT is included in both fits.  The first also
includes a multicolor disk-blackbody (DBB), while the second includes
a blackbody (BB).  Values are in keV unless   
otherwise noted (f and $\tau_p$ are unitless). Values from
normalizations utilize the 
PCA normalizations; uncertainty in PCA absolute normalization is not
included.}
\tablenotetext{a}{kT and R refer to the inner
edge of the diskbb for model 1 (where R is R$_{in}$cos(i)$^{0.5}$),
and to the blackbody for model 2.} 
\tablenotetext{b}{f is the ratio of the blackbody or diskbb flux to
COMPTT flux, over 0.1-100 keV range.}

\end{deluxetable}

\begin{deluxetable}{lr}
\tablewidth{2.0truein}
\tablecaption{\textbf{Spectral Fits to LMXB EXO 1745-248 Features}}
\tablehead{\multicolumn{2}{c}{Gaussian}
}
\startdata
kT & 6.55$^{+.06}_{-.07}$ \\
EW (eV) & 188$^{+86}_{-83}$ \\
$\sigma$ & 0.31$^{+.16}_{-.15}$ \\
\hline
\multicolumn{2}{c}{Smeared edge} \\
\hline
kT & 8.1$\pm0.8$ \\
$\tau_{\rm max}$ & 0.25$^{+1.22}_{-.16}$ \\
Width & 2.6$^{+5.5}_{-2.0}$  \\
\enddata
\tablecomments{Spectral fits to the Fe-line gaussian and disk
reflection smeared edge in the spectrum of the LMXB
EXO 1745-248.  The spectral fits in this table are to the ``standard
model'' including a multicolor disk-blackbody (see text); line and
edge parameter values for the other
model considered in the text are within the errors in this table.
Values are in keV 
unless otherwise stated, except $\tau_{\rm max}$ which is unitless.
}

\end{deluxetable}

\clearpage


\psfig{file=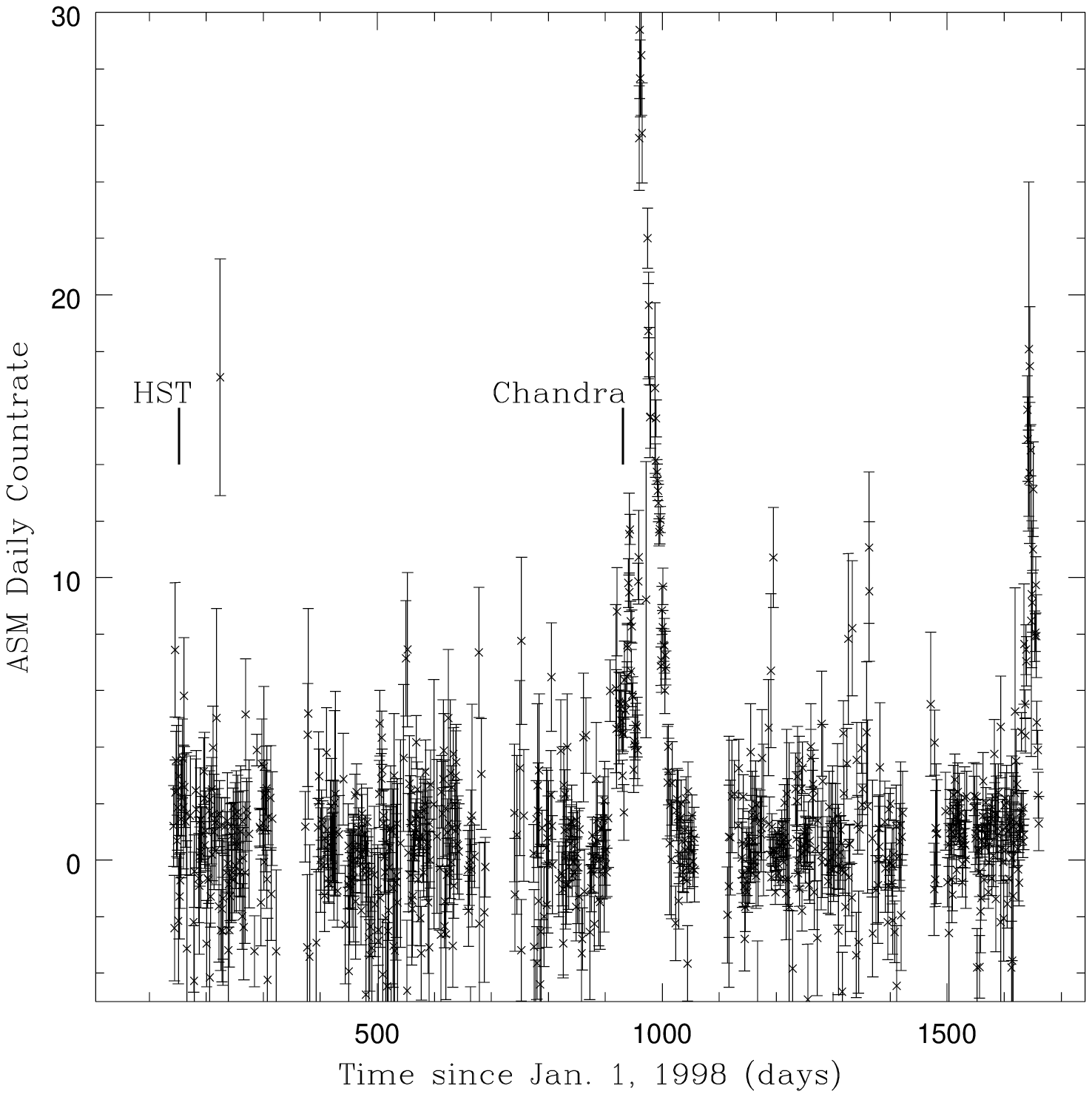}

\figcaption[f1.eps]{The \RXTE All-Sky Monitor lightcurve of the
LMXB EXO1748-25 in Terzan 5.  The dates of the \Chandra and \HST 
observations are marked.
\label{Figure 1}}

\psfig{file=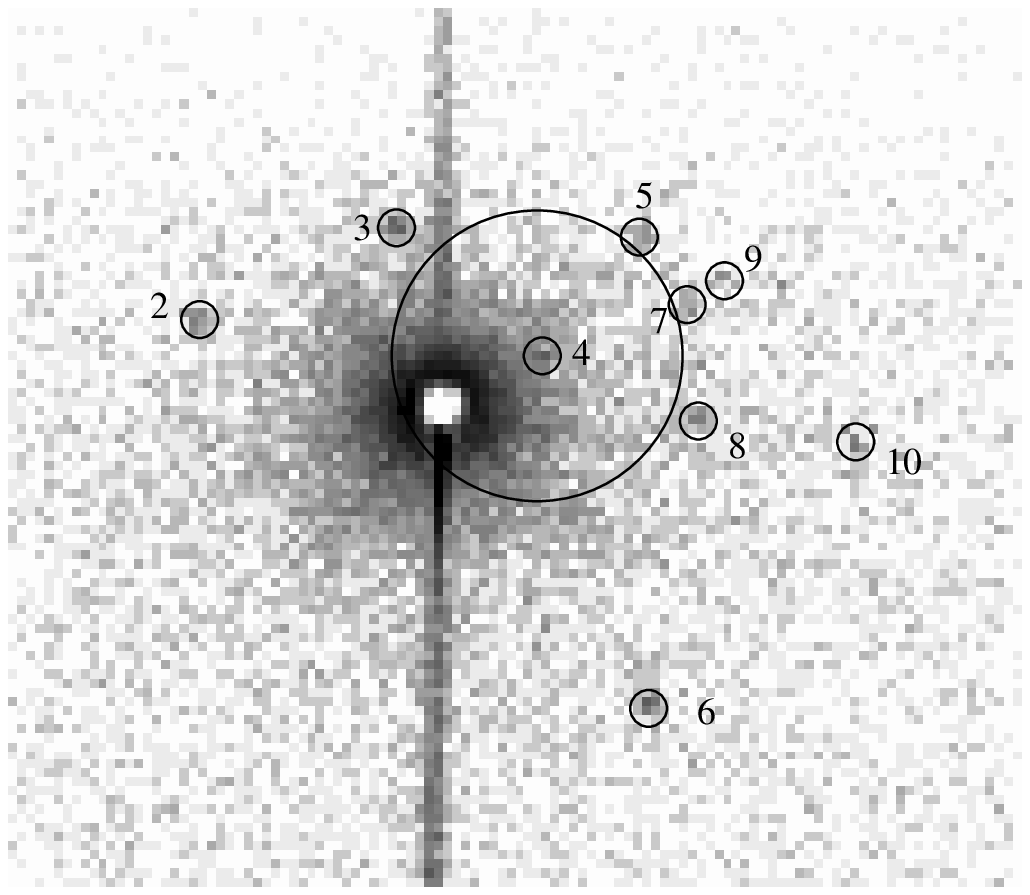}

\figcaption[f2.eps]{\Chandra ACIS-I image of the globular
cluster Terzan 5, energy range 0.5-2.0 keV.  The dominant feature is
the piled-up halo from the LMXB EXO 1745-248 in its high state.  The
streak is due to 
out-of-time LMXB events recorded during the frame transfer, and the
position of the LMXB has no good events due to pulse saturation.  Several
other sources are visible, marked with 1'' circles and indicated with
their shorthand names.  The cluster core is indicated with a 7.9'' (1
$r_c$) circle.
\label{Figure 2}}

\psfig{file=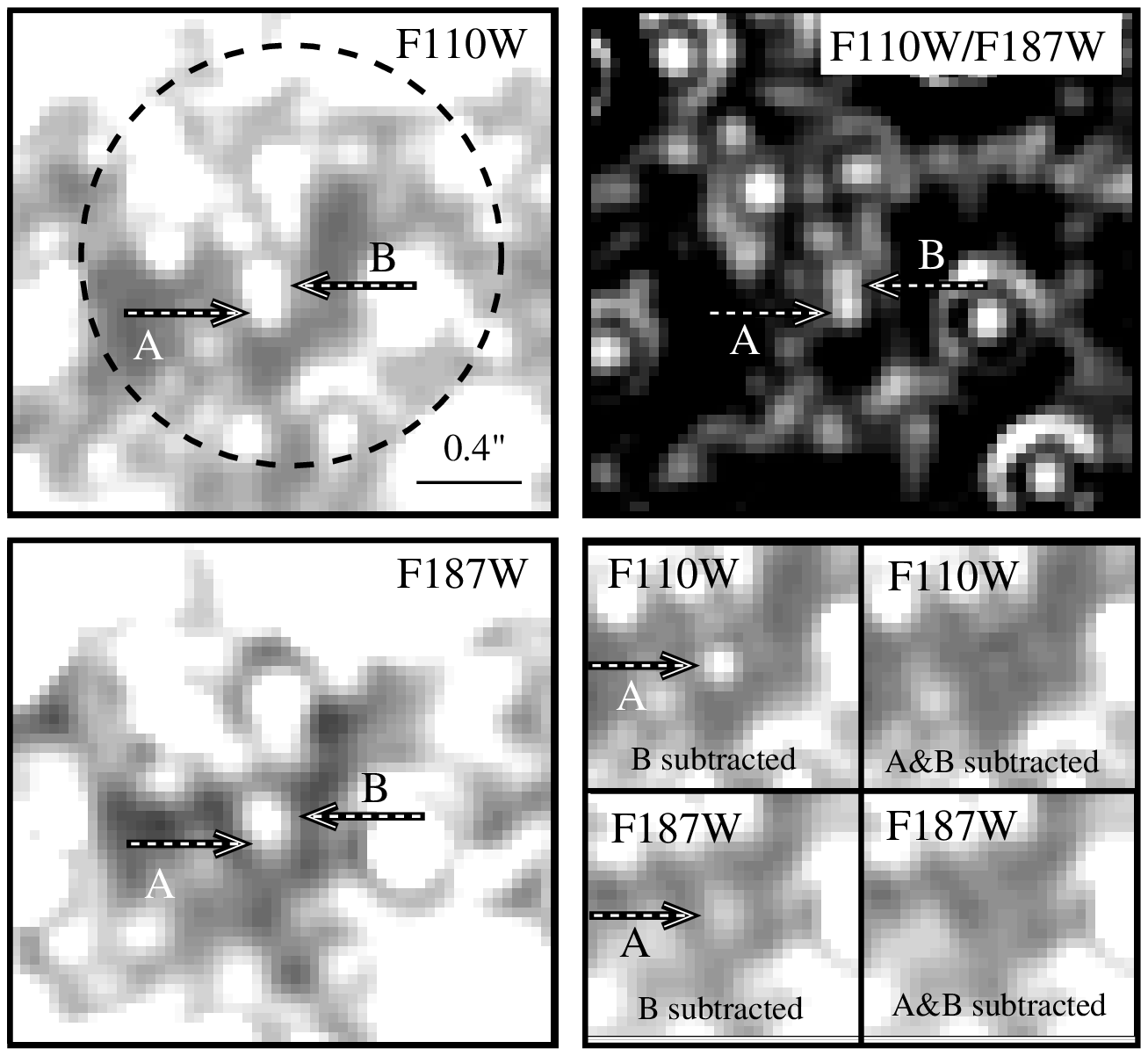}
\figcaption[f3.eps]{Finding chart, showing stars A and B in F110W
(upper left)
and F187W (lower left) \HST NICMOS data, and the ratio image F110W/F187W
(upper right).  The best
estimate of the uncertainty in the position of EXO 1745-248 is indicated by the
0.8'' (2 $\sigma$) error circle.    Note that star
A is the bluest object in this field (some red giant cores appear blue
due to saturation).  The results of our manual psf
fitting of stars A and B are shown at lower right.  \label{Figure 3}}

\psfig{file=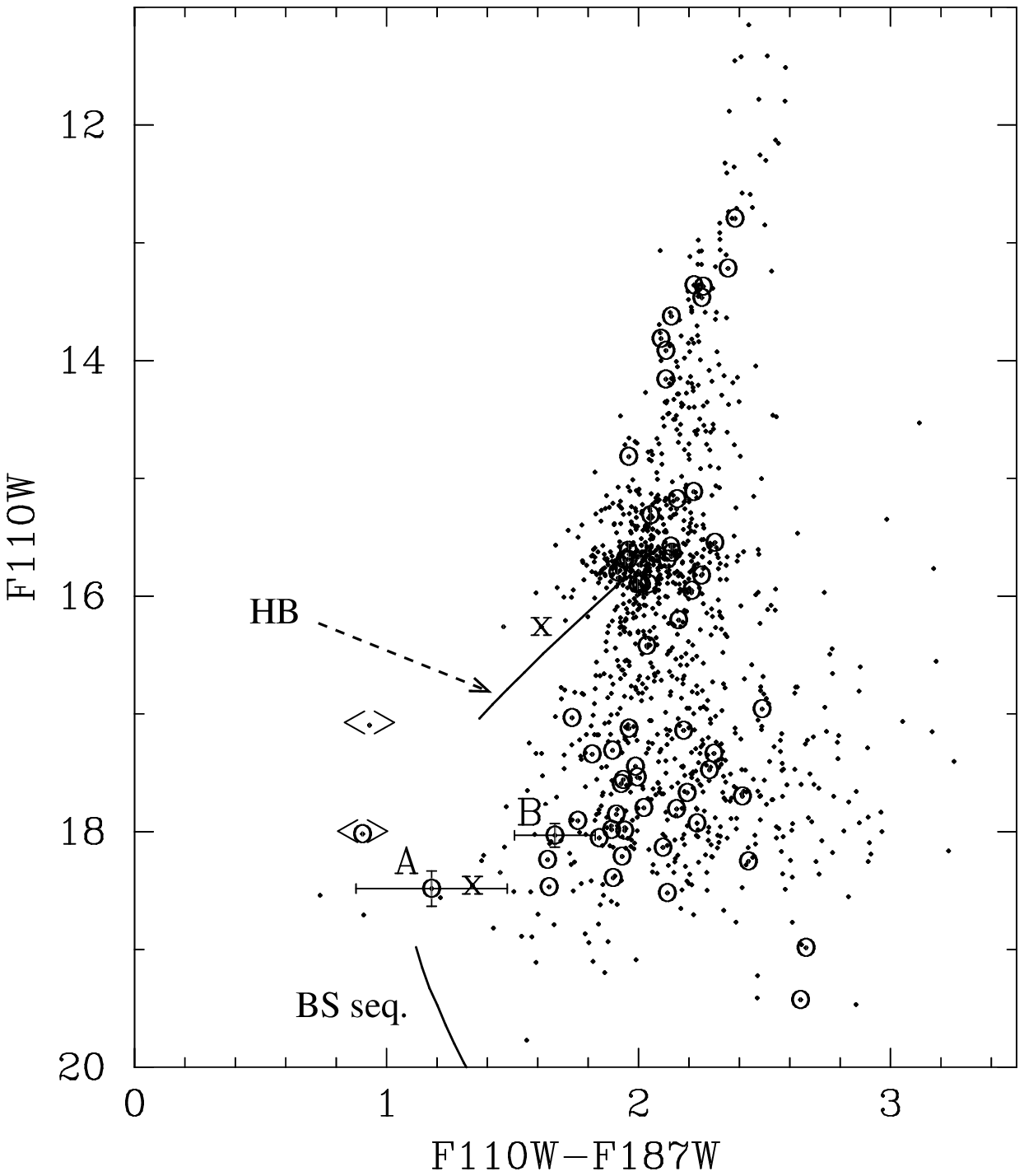}
\figcaption[f4.eps]{Aperture photometry color-magnitude diagram
of Terzan 5, from the \HST NIC2 camera (using highly dithered imaging via
the drizzle algorithm) on the core of Terzan 5.  Stars from
the photometry of EGC01 
within 1'' of the seven X-ray error circles are circled.  Xs mark
the locations of variable stars from EGC01 (X2 is the dimmer one).  Two stars
that are more than 3 $\sigma$ to the blue of the distribution are
indicated by $<.>$.  The lower is the blend of stars A and B (see
text).   The separately derived positions of 
stars A and B are indicated, with errors from the photometry.  The
horizontal branch and the expected blue straggler sequence,
terminating in a Kurucz model (Kurucz 1992) of a 1.6 \Msun\ star (twice
the turnoff mass)  are indicated.  
\label{Figure 4}}

\psfig{file=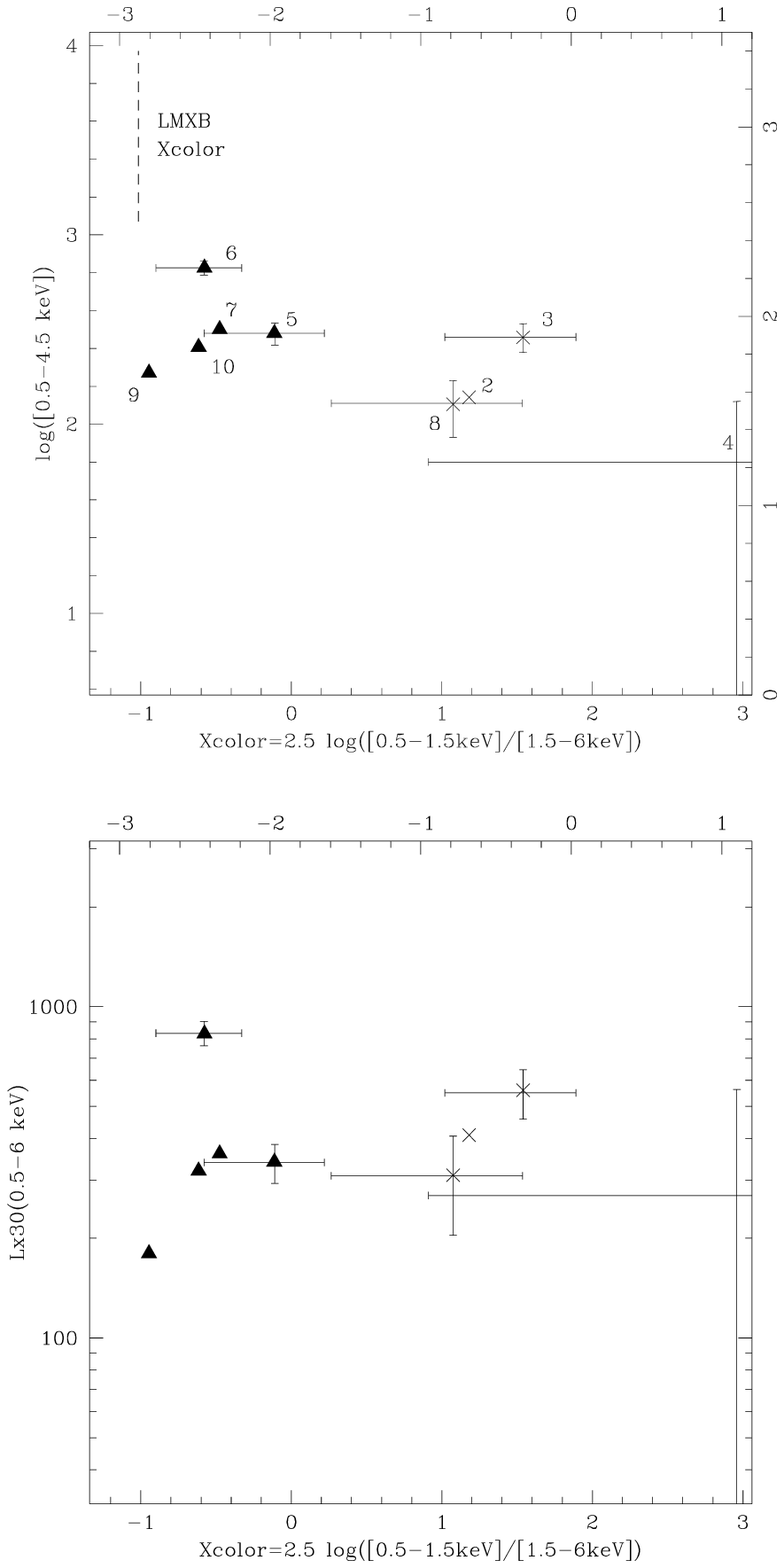}

\figcaption[f5.eps]{X-ray color-magnitude diagrams of Terzan 5. Upper
panel 
produced from the absorption-corrected counts in the 0.5-1.5, 0.5-4.5,
and 1.5-6 
keV energy bands (lower and left axes; upper and right axes provide
the observed colors and magnitudes).  We correct roughly for
photoelectric absorption by shifting the data +0.57 units on the left
axis and +1.86 units on the bottom axis.  However, note that the
effects of absorption are not uniform.  The
probable qLMXBs are marked by crosses, and
probable CVs by filled triangles. We plot representative errors for a
few points.  Background subtraction leaves W4
with -1 counts in the hard band; we show its Xcolor lower limit and
counts range (it is only clearly detected in the soft band).  The Xcolor
of EXO 1745-248's X-ray halo is indicated with a
dotted line and labeled 'LMXB Xcolor'. 
Lower panel produced using the same absorption-corrected color for x axis,
but using the unabsorbed 0.5-6 keV luminosity for y axis.  Luminosities
derived from best-fitting hydrogen atmosphere fits for W2, W3, W4, and
W8, and best-fitting thermal bremsstrahlung fit for W5, W6, W7, W9,
and W10.  Luminosity errors are based on 0.5-4.5 keV counting
statistics, and do not include uncertainties in spectral fitting.
\label{Figure 5}}

\psfig{file=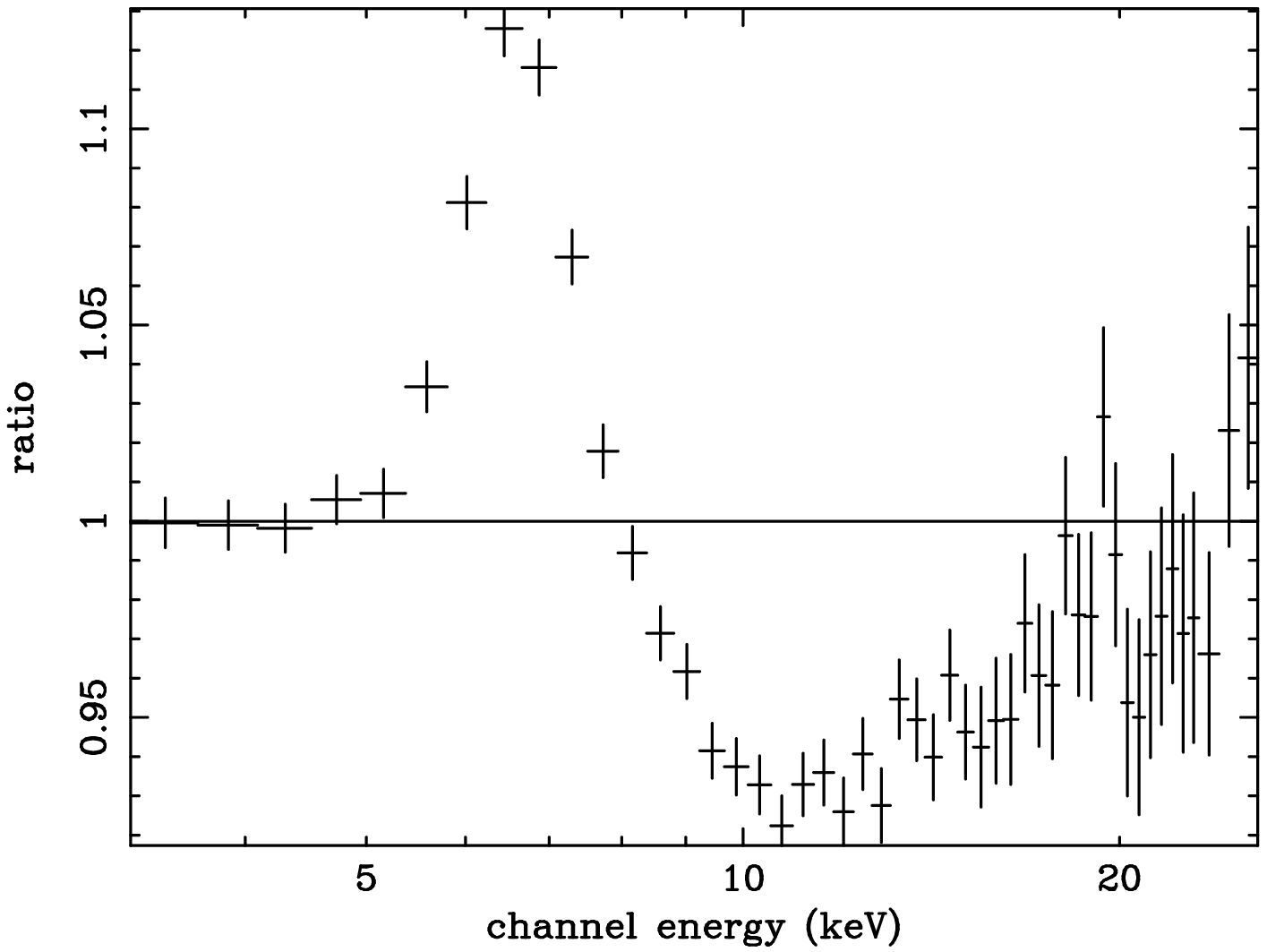}

\figcaption[f6.eps]{Ratio of EXO 1745-248 PCA data to standard model
(see text), with
normalizations of the gaussian iron line and smeared edge set to zero
to show the relative contributions.
\label{Figure 6}}

\psfig{file=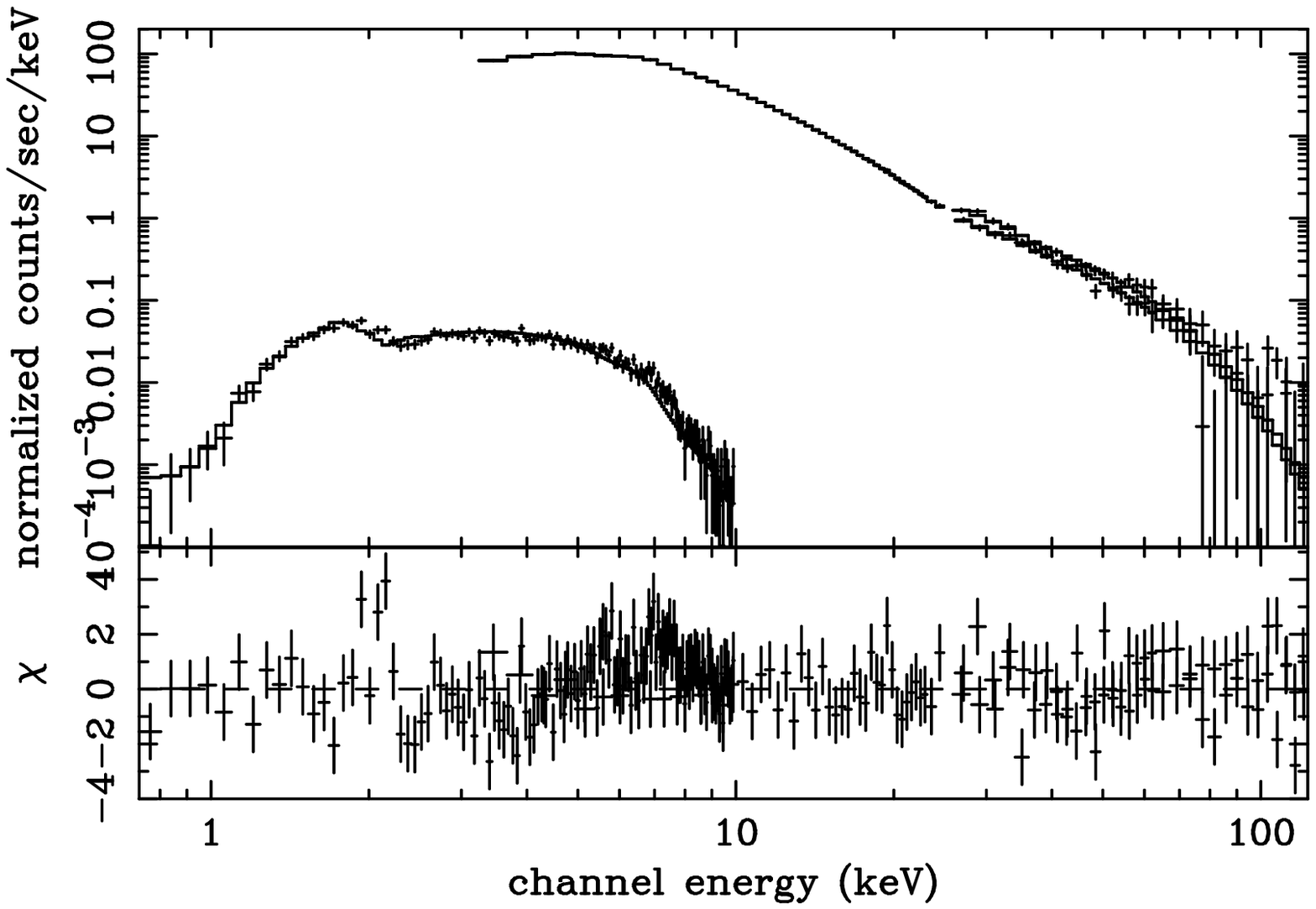}

\figcaption[f7.eps]{EXO 1745-248 spectrum from \Chandra readout streak
data, \RXTE PCA, and \RXTE HEXTE data, simultaneously fitted with the
standard model 
(see text). Note the poor fit to the \Chandra data near 2 keV and the
pronounced wave above 5 keV, suggesting an incorrect energy calibration.
\label{Figure 7}}

\psfig{file=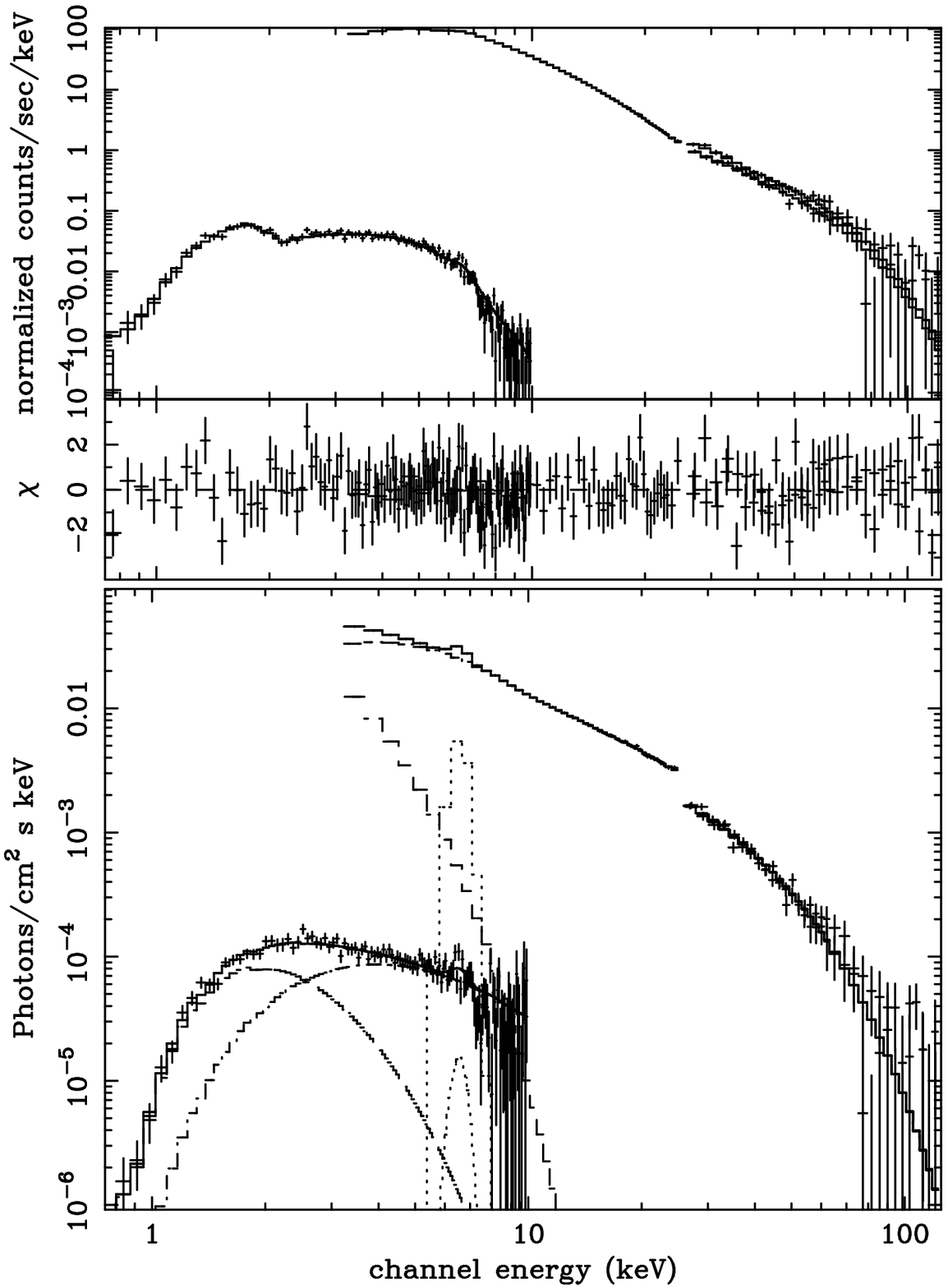}

\figcaption[f8.eps]{EXO 1745-248 observed (top) and unfolded (bottom)
spectra, with best-fit absorbed 
(DISKBB+COMPTT+GAUSSIAN)$\times$SMEDGE model spectrum.  Here the
energies assigned to \Chandra readout streak events have been
increased by 7\%.   Relative normalizations of the different
instruments are allowed to vary (see text). The 6.55
keV iron line is visible in both the \RXTE and \Chandra unfolded
spectra, along with the DISKBB and COMPTT components (the COMPTT is
the harder component). 
\label{Figure 8}}


\begin{thebibliography}{33}
\expandafter\ifx\csname natexlab\endcsname\relax\def\natexlab#1{#1}\fi

\bibitem[{Aldcroft} {et~al.}(2000)]{Aldcroft2000}
{Aldcroft}, T.~L., {Karovska}, M.,
{Cresitello-Ditmar}, M.~L., {Cameron}, R.~A., {Markevitch},
M.~L. 2000, Proc. SPIE, 4012, 650
 
\bibitem[{{Arnaud} (1996)}]{Arnaud96}
{Arnaud}, K.~A. 1996, in G. Jacoby \& J. Barnes, (eds.) ASP
Conf. Series {\it Astronomical Data Analysis Software and
Systems V.}, vol. 101, 17

\bibitem[Asai et al. (2000)]{Asai00}
Asai, K., Dotani, T., Nagase, F., \& Mitsuda, K. 2000, ApJS 131, 571

\bibitem[Barret et al. (2000)]{Barret00}
Barret, D., Olive, J.~F., Boirin, L., Done, C., Skinner, G.~K., \&
Grindlay, J.~E. 2000, ApJ 533, 329

\bibitem[Becker et al. (2003)]{Becker03}
Becker, W. et al.~2003, ApJ submitted (astro-ph/0211468)

\bibitem[Bloser et~al. (2000)]{Bloser00}
Bloser, P.~F., Grindlay, J.~E., Kaaret, P., Zhang, W., Smale, A.~P.,
\& Barret, D. 2000, ApJ 542, 1000

\bibitem[Cohn et~al.(2002)]{Cohn02}
Cohn, H.~N., Lugger, P.~M., Grindlay, J.~E., \& Edmonds, P.~D. 2002,
ApJ 571, 818

\bibitem[Cool et al.(2002)]{Cool02}
Cool, A.~M., Haggard, D., Carlin, J.~L. 2002, in F. van Leeuwen,
J.~D. Hughes, G. Piotto (eds.), Omega Centauri, A Unique Window into
Astrophysics, Vol. 265 of {\it ASP Conference Series}, ASP, p. 277

\bibitem[D'Amico et al.~(2002)]{D'Amico02}
D'Amico, N., Possenti, A., Fici, L., Manchester, R.~N., Lyne, A.~G.,
Camilo, F., \& Sarkissian, J. 2002, ApJ 570, L89

\bibitem[Daniel (1990)]{Daniel90}
Daniel, W.~W. 1990, {\it Applied Nonparametric Statistics, 2d ed.}, PWS-Kent

\bibitem[Danner et al. (1997)]{Danner97}
Danner, R., Kulkarni, S.~R., Saito, Y., \& Kawai, N. 1997, {\it
Nature} 388, 751

\bibitem[Davies \& Hansen (1998)]{Davies98}
Davies, M.~B., \& Hansen, B.~M.~S. 1998, MNRAS 301, 15

\bibitem[{{Deutsch} {et~al.}(2000){Deutsch}, {Margon}, \& {Anderson}}]{DMA00}
{Deutsch}, E.~W., {Margon}, B., \& {Anderson}, S.~F. 2000, \apj, 530,
L21

\bibitem[Djorgovski (1993)]{Djor93}
Djorgovski, S. 1993, in S. Djorgovski, G. Meylan (eds.), Structure and
Dynamics of Globular Clusters, Vol. 50 of {\it ASP Conference Series},
ASP, p. 373

\bibitem[Ebisawa et~al. (1994)]{Ebisawa94}
Ebisawa, K. et al.~1994, PASJ 46, 375

\bibitem[Edmonds et~al.(2001)]{Edmonds01}
Edmonds, P.~D., Grindlay, J.~E., Cohn, H.~N.,  \& Lugger, P.~M. 2001,
ApJ 547, 829

\bibitem[Fruchter \& Goss (2000)]{Fruchter00}
Fruchter, A.~S. \& Goss, W.~M. 2000, ApJ 536, 865

\bibitem[Gehrels (1986)]{Gehrels86}
Gehrels, N. 1986, ApJ 303, 336

\bibitem[Giacconi et~al. (2001)]{Giacconi01}
Giacconi, R. et al.~2001, ApJ 551, 624

\bibitem[{{Grindlay} {et~al.}(2001){Grindlay}, {Heinke}, {Edmonds}, \&
{Murray}}]{Grindlay01a}
{Grindlay}, J.~E., {Heinke}, C.~O., {Edmonds}, P.~D., \& {Murray},
S.~S. 2001a,  {\it Science} 292, 2290 

\bibitem[{{Grindlay} {et~al.}(2001){Grindlay}, {Heinke}, {Edmonds},
{Murray}, \& {Cool}}]{Grindlay01b}
{Grindlay}, J.~E., {Heinke}, C.~O., {Edmonds}, P.~D., {Murray},
S.~S., \& Cool, A.~M. 2001b,  ApJ 563, L53

\bibitem[Grindlay et al. (2002)]{Grindlay02}
Grindlay, J.~E., Camilo, F., Heinke, C.~O., Edmonds, P.~D., Cohn, H.,
\& Lugger, P. 2002, ApJ in press (available at astro-ph/0208280)

\bibitem[Grindlay et al. (2003)]{Grindlay03}
Grindlay, J.~E. et al.~2003, AN in press (available at astro-ph/0211527)

\bibitem[Harris (1996)]{Harris96}
Harris, W.~E. 1996, AJ, 112, 1487

\bibitem[Heinke et~al. (2001)]{Heinke01}
Heinke, C.~O., Edmonds, P.~D., Grindlay, J.~E. 2001, ApJ 562, 363

\bibitem[Heinke et~al. (2002)]{Heinke02}
Heinke, C.~O., Grindlay, J.~E., Lloyd, D.~A., \& Edmonds, P.~D. 2002,
ApJ, submitted 

\bibitem[Hertz \& Grindlay (1984)]{Hertz84}
Hertz, P. \& Grindlay, J.~E. 1984, ApJ 282, 118

\bibitem[H\"unsch et al. (1999)]{Hunsch99}
H\"unsch, M., Schmitt, J.~H.~M.~M, Sterzik, M.~F., \& Voges, W. 1999, A\&AS,
135, 319

\bibitem[Inoue et~al. (1984)]{Inoue84}
Inoue, H. et al.~ 1984, PASJ 36, 855

\bibitem[Johnston et~al. (1995)]{Johnston95}
Johnston, H.~M., Verbunt, F., \& Hasinger, G. 1995, A\&A 298, L21

\bibitem[Johnston \& Verbunt (1996)]{Johnston96}
Johnston, H.~M. \& Verbunt, F. 1996, A\&A 312, 80

\bibitem[Jonker et al.~(2003)]{Jonker03}
Jonker, P.~G., Mendez, M., Nelemans, G., Wijnands, R., \& van der
Klis, M 2003, MNRAS (in press; astro-ph/0301475)

\bibitem[Kurucz (1992)]{Kurucz92}
Kurucz, R. 1992, Rev.Mex.A.A. 23, 181

\bibitem[Kuulkers et al. (2002)]{Kuulkers02}
Kuulkers, E., den Hartog, P.~R., in't Zand, J.~J.~M., Verbunt,
F.~W.~M., Harris, W.~E., \& Cocchi, M. 2002, A\&A (in press; astro-ph/0212028)


\bibitem[Lloyd (2003)]{Lloyd03}
Lloyd, D.~A. 2003, MNRAS (submitted)

\bibitem[Lugger, Cohn, \& Grindlay (1996)]{Lugger96}
Lugger, P.~M., Cohn, H.~N., \& Grindlay, J.~E. 1996, ApJ 439, 191

\bibitem[Lyne et~al. (2000)]{Lyne00}
Lyne, A.~G., Mankelow, S.~H., Bell, J.~F., \& Manchester, R.~N. 2000,
MNRAS, 316, 491

\bibitem[Makishima et~al. (1981)]{Makishima81}
Makishima, K. et al.~1981, ApJ 247, L23

\bibitem[Makishima et~al. (2000)]{Makishima00}
Makishima, K. et al.~2000, ApJ 535, 632

\bibitem[Markwardt \& Swank (2000a)]{Mark00a}
Markwardt, C.~B. \& Swank, J.~H. 2000a, IAU Circ. 7454

\bibitem[Markwardt et al. (2000b)]{Mark00b}
Markwardt, C.~B., Strohmayer, T.~E., Swank, J.~H., \& Zhang, W. 2000b,
IAU Circ. 7482 

\bibitem[Merloni et al. (2000)]{Merloni00}
Merloni, A., Fabian, A.~C., \& Ross, R.~R. 2000, MNRAS 313, 193

\bibitem[Ortolani et al. (1996)]{Ortolani96}
Ortolani, S., Barbuy, B., \& Bica, E. 1996, A\&A 308, 733

\bibitem[Ortolani et al. (2001)]{Ortolani01}
Ortolani, S., Barbuy, B., Bica, E., Renzini, A., Zoccali, M., Rich,
R.~M., \& Cassisi, S., \aap, 2001, 376, 878

\bibitem[Parmar et al.~(2001)]{Parmar01}
Parmar, A.~N., Oosterbroek, T., Sidoli, L, Stella, L., \& Frontera,
F. 2001, A\&A 380, 490

\bibitem[Pooley et~al. (2002a)]{Pooley01a}
Pooley, D., et~al. 2002, ApJ 569, 405

\bibitem[Pooley et~al. (2002b)]{Pooley01b}
Pooley, D., et~al. 2002, ApJ  573, 184

\bibitem[{{Predehl} \& {Schmitt}(1995)}]{predehl95}
Predehl, P. \& Schmitt, J.~H.~M.~M. 1995, A\&A 293, 889

\bibitem[{{Rutledge} {et~al.}(2001c)}]{Rutledge01c}
{Rutledge}, R.~E., {Bildsten}, L., {Brown}, E.~F., {Pavlov},
G.~G., \& {Zavlin}, V.~E. 2001c,  ApJ 578, 405

\bibitem[Shimura \& Takahara (1995)]{Shimura95}
Shimura, T., \& Takahara, F. 1995, ApJ 445, 780

\bibitem[Sidoli et al. (2001)]{Sidoli01}
Sidoli, L, Parmar, A.~N., Oosterbroek, T., Stella, L., Verbunt, F.,
Masetti, N., \& Dal Fiume, D. 2001, A\&A 368, 451

\bibitem[Smale (1995)]{Smale95}
Smale, A.~P. 1995 AJ, 110, 1292

\bibitem[Smith et al.~(2002)]{Smith02}
Smith, R.~K., Edgar, R.~J., \& Shafer, R.~A. 2002, ApJ 581, 562

\bibitem[Titarchuk (1994)]{Titarchuk94}
Titarchuk, L. 1994, ApJ 434, 570

\bibitem[Trager et al. (1995)]{Trager95}
Trager, S.~C., King, I.~R., \& Djorgovski, S. 1995, AJ, 109, 218

\bibitem[{{van Paradijs} \& {McClintock}(1994)}]{van Paradijs1994}
{van Paradijs}, J. \& {McClintock}, J.~E. 1994, \aap, 290,
133

\bibitem[{{van Paradijs} \& {McClintock}(1995)}]{van Paradijs1995}
{van Paradijs}, J. \& {McClintock}, J.~E. 1995, in X-ray Binaries,
ed. Lewin, van Paradijs, van den Heuvel (Cambridge U. Press), p. 58

\bibitem[Verbunt \& Hut (1987)]{Verbunt87}
Verbunt, F., \& Hut, P. 1987, IAU Symp. 125, 187

\bibitem[{{Verbunt} {et~al.}(1995){Verbunt}, {Bunk}, {Hasinger}, \&
{Johnston}}]{Verbunt95}   
{Verbunt}, F., {Bunk}, W., {Hasinger}, G. \& {Johnston}, H.~M., 1995, 
\aap, 300, 732


\bibitem[Verbunt (2002)]{Verbunt02}
Verbunt, F. 2002, in ASP Conf. Ser. {\it $\omega$ Centauri, a unique window in
astrophysics}, ed. van Leeuwen, Piotto, Hughes (available at astro-ph/0111441)

\bibitem[Wachter (1997)]{Wachter97}
Wachter, S. 1997, ApJ 485, 839

\bibitem[Wachter (1998)]{Wachter98}
Wachter, S. 1998, Ph.D. thesis, University of Washington

\bibitem[Wang et~al. (2001)]{Wang01}
Wang, Z. et~al. 2001, ApJ 563, L61

\end{thebibliography}
\end{document}